\documentclass[aps,pra,twocolumn,superscriptaddress]{revtex4-2}

\usepackage{amsfonts,mathrsfs,amsmath,amsthm,amssymb}
\usepackage{bm,times}
\usepackage{graphicx,epsfig}
\usepackage[colorlinks=true,breaklinks=true,linkcolor=blue,citecolor=blue,urlcolor=blue]{hyperref}

\def \tr{{\rm{Tr}}}
\def \non {\nonumber}

\def \ket {\rangle}
\def \bra {\langle}

\newcommand{\be}{\begin{eqnarray}}
\newcommand{\ee}{\end{eqnarray}}

\makeatletter
    
    \newcommand{\Rmnum}[1]{\expandafter\@slowromancap\romannumeral #1@}
\makeatother

\begin{document}
\title{Averaged fidelity-based steering criteria}
\author{Xiaohua Wu}
\email{wxhscu@scu.edu.cn}
\affiliation{College of Physical Science and Technology, Sichuan University, Chengdu 610064, China}

\author{Bo You}
\affiliation{College of Physical Science and Technology, Sichuan University, Chengdu 610064, China}

\author{Tao Zhou}
\email{taozhou@swjtu.edu.cn}
\affiliation{Quantum Optoelectronics Laboratory, School of Physical Science and Technology, Southwest Jiaotong University, Chengdu 610031, China}
\affiliation{Department of Physics, School of Physical Science and Technology, Southwest Jiaotong University, Chengdu 611756, China}

\begin{abstract}
In the present work, the averaged fidelity is introduced as the steering parameter. According to the definitions of steering from Alice to Bob, a general scheme for designing linear steering criteria is developed for a high-dimensional system. For a given set of measurements on Bob's side, two quantities, the so-called nonsteering thresholds, can be defined. If the measured averaged fidelity exceeds these thresholds, the state shared by Alice and Bob is steerable from Alice to Bob, and the measurements performed by Alice are also verified to be incompatible. Within the general scheme, we also construct a linear steering inequality when the set of measurements performed by Bob has a continuous setting. Some applications are also provided.
\end{abstract}

%\pacs{03.65.Ud, 03.65.Ta}
%03.65.Ud Entanglement and quantum nonlocality
%03.65.Ta Foundations of quantum mechanics; measurement theory

\date{\today}

\maketitle

\section{Introduction}
The concept of steering can date back to 1930s, introduced by Schr\"odinger~\cite{Sch} as a generalization of the Einstein-Podolsky-Rosen (EPR) paradox~\cite{Ein}. For a bipartite state, steering infers that an observer on one side can affect the state of the other spatially separated system by local measurements. In quantum information processing, steering can be defined as the task for a referee to determine whether two parties share entanglement with an untrusted party~\cite{Wiseman1,JWD,sau}. In 2007, Wiseman, Jones and Doherty~\cite{Wiseman1} formally defined quantum steering  as a type of quantum nonlocality that is logically distinct from inseparability~\cite{Guhne,Horos} and Bell nonlocality~\cite{Brunner}. In the modern view, quantum steering can be understood as the impossibility of describing the conditional states at one party by a local hidden state (LHS) model.

A fundamental property is that steering is inherently asymmetric with respect to the observers~\cite{bowles,Midgley}, which is quite different from the quantum nonlocality and entanglement. Actually, there are entangled states which are one-way steerable~\cite{bowles,Bow}. Besides its foundational significance in quantum information theory, steering has been found useful in many applications. For examples, steering has a vast range of information-theoretic applications in one-sided device-independent scenarios where the party being steered has trust on his or her own quantum device while the other's device is untrusted, such as one-sided device-independent quantum key distribution~\cite{Bran}, advantage in subchannel discrimination \cite{piani}, secure quantum teleportation \cite{Reid1,He}, quantum communication~\cite{Reid1}, detecting bound entanglement \cite{Mor}, one-sided device-independent randomness generation~\cite{law}, and one-sided device-independent self-testing of pure maximally as well as nonmaximally entangled state~\cite{supic}.

Meanwhile, the detection and characterization of steering, especially the steering inequlities, have been widely discussed. In 1989, the variance inequalities violated with EPR correlations for the continuous variable system were derived by Reid~\cite{eid}, and this was generalized to discrete variable systems~\cite{Caval}. EPR-steering inequalities were defined~\cite{can22}, where the violation of any such inequality implies steering. Following these works, further schemes have been proposed to signalize steering, for instance, the linear and nonlinear steering criteria~\cite{sau,wit,Pusey,Evan,mar,rut}, steering inequalities based on multiplicative variances~\cite{ReidRMD}, steering criteria from uncertainty relations~\cite{wa,schnee,Costaa,costab,jia,kri}, steering with Clauser-Horne-Shimony-Holt (CHSH)-like inequalities~\cite{Can3,Girdhar,cos,quan}, moment matrix approach~\cite{Kig,mo,chen00}, steering criteria based on local uncertainty relations~\cite{Ji,Zhen}, and the universal steering criteria~\cite{Zhu}. The discussed criteria or small variation thereof have been used in several experiments~\cite{sau,wit,Bennet,smith,weston}. Quite recently, the connection between quantum steering and quantum coherence was discussed~\cite{Mondal,Mondal1}.

Our  work is originated from one of the open questions summarized in Ref.~\cite{rmd}: Though a complete characterization of quantum steerability has been obtained for two-qubit systems and projective measurements, it is still desirable to extend such a characterization  to higher-dimensional systems. Although there is indication that such an extension is possible, much remains to be worked out. Here, we shall focus on finding the  sufficient criteria for steerability with   linearly steering inequalities  (LSIs) \cite{can22, sau,Joness}. The LSIs have an advantage that they can work even when the bipartite state is unknown. They also have a deep relation with the joint measurement problem: If an LSI is violated, the state is steerable from Alice to Bob and the measurements performed by Alice are also verified to be incompatible~\cite{quint,ula,UULA,Kiukas,Wu}.

In this paper, according to the definitions of steering~\cite{Can1}, we develop a general scheme to design linear steering criteria for high-dimensional systems by introducing the averaged fidelity as the steering parameter. The  content of this work is organized as follows. In Sec.~\ref{Sec2}, we give a brief review on the definition of steering from Alice to Bob, the Werner states and the isotropic states. In Sec.~\ref{Sec3}, a detailed introduction to the non-steering threshold is given. In Sec.~\ref{Sec4}, we address the problem of constructing linear criteria for high-dimensional systems, and an explicit LSI is constructed. Some applications of the LSI are discussed in Sec.~\ref{Sec5}. Finally, we end our work with a short conclusion.

\section{Preliminary}
\label{Sec2}
\subsection{Steering from Alice to Bob}

A bipartite state $W$ shared by Alice and Bob can be expressed by a pure (entangled) state and a one-sided linear map $\varepsilon$~\cite{horo1,Ruskai},
\begin{equation}
\label{decoposition}
W=\mathbb{I}_d\otimes \varepsilon(\vert \Psi\rangle\langle \Psi\vert),
\end{equation}
where $\mathbb{I}_d$ is an identity map. Let $\rho_{\mathrm{A}}$ be the reduced density matrix on Alice's side, and $\vert\Psi\rangle$ could be fixed as $\vert\Psi\rangle=\vert \sqrt{\rho_{\mathrm{A}}}\rangle \rangle$. The details about $\varepsilon$ and $\vert \sqrt{\rho_{\mathrm{A}}}\rangle \rangle$ are shown in Appendix~\ref{appA}. Usually, $\rho_{\mathrm{A}}$ has an eigen decomposition $\rho_{\mathrm{A}}=\sum_{i=1}^d\lambda_i\vert i\rangle\langle i\vert$, with $d$ the dimension of the Hilbert space, and $\vert\Psi\rangle=\sum _{i=1}^d \sqrt{\lambda_i}\vert i\rangle\otimes \vert i\rangle.$ In the field of quantum information and computation, entanglement is one of the most important quantum resources, and it is important to verify whether a bipartite state $W$ is entangled or not. Based on this decomposition, the state $W$ should be a mixture of products states  if and only if $\varepsilon$ is entanglement-breaking (EB)~\cite{horo1,Ruskai}.

Before one can show how to demonstrate a state is steerable from Alice to Bob, some necessary conventions are required. First, Alice can perform $N$ projective measurements on her side, labeled by $\mu=1,2,...,N$, each having $d$ outcomes $a=0,1,...,d-1$, and the measurements are denoted by $\hat{\Pi}^{a}_{\mu}$, $\sum_{a=1}^{d}\hat{\Pi}^{a}_{\mu}=I_d$, with $I_d$ the identity operator on the $d$-dimensional Hilbert space. The unnormalized postmeasurement states for Bob are
\begin{equation}
\label{df}
\tilde{\rho}_{\mu}^a=\mathrm{Tr}_\mathrm{A}\left[(\hat{\Pi}^{a}_{\mu}\otimes I_d)W\right],
\end{equation}
and from the decomposition in Eq.~\eqref{decoposition}, it can be rewritten as
\begin{equation}
\label{unnormalized}
\tilde{\rho}_{\mu}^a=\varepsilon\left(\sqrt{\rho_{\mathrm{A}}}(\hat{\Pi}_{\mu}^a)^*\sqrt{\rho_{\mathrm{A}}}\right).
\end{equation}
where ``$*$'' represents the complex conjugate.
 The set of unnormalized states, $\{\tilde{\rho}^{a}_{\mu}\}$, is usually called an \emph{assemblage}.

In 2007, Wiseman , Jones and Doherty formally defined quantum steering as the possibility of remotely generating ensembles that could not
be produced by \emph{a local hidden states} (LHS) model  ~\cite{Wiseman1}. A LHS model refers to the case where a source sends a classical message $\xi$ to one of the two parties, say, Alice, and a corresponding quantum state $\rho_{\xi}$ to the other party, say Bob.   Given that  Alice decides to perform the $\mu$-th measurement, the variable $\xi$ instructs the output $a$ of Alice's apparatus with the probability $\mathfrak{p}(a\vert\mu,\xi)$. The variable $\xi$ also can be interpreted as a local-hidden-variable (LHV) and  chosen according to a probability distribution $\Omega(\xi)$.
  Bob does not have access to the classical variable $\xi$, and his final assemblage is composed by
\begin{equation}
\label{tilderho}
\tilde{\rho}^{a}_{\mu}=\int d\xi\Omega(\xi)\mathfrak{p}(a\vert\mu,\xi)\rho_{\xi},
\end{equation}
Note that $\tr\tilde{\rho}^a_{\mu}$ is the probability that the outcome is $a$ when the $\mu$-th measurement is performed by Alice, and in the LHS model above,
there is $\mathrm{Tr}[\tilde{\rho}^a_{\mu}]=\int d\xi\Omega(\xi)\mathfrak{p}(a\vert\mu,\xi)$.

In this paper, the definition of steering is directly cited from the review article~\cite{Can1}: An assemblage is said to demonstrate steering if it does not admit a decomposition of the form in Eq.~\eqref{tilderho}.  Furthermore, a quantum state $W$ is said to be steerable from Alice to Bob if the experiments in Alice's part produce an assemblage that demonstrates steering. On the contrary, an assemblage is said to be LHS if it can be written as in Eq.~\eqref{tilderho}, and a quantum state is said to be unsteerable  if an LHS assemblage is generated for all local measurements.
 
For a given class of bipartite states, how to construct the necessary and sufficient condition of steerability is one of the elementary tasks in quantum steering. Until now, the necessary and sufficient criteria have been obtained only for the Werner states~\cite{Werner}, isotropic states~\cite{Wiseman1}, and $T$-states~\cite{jev}.  In the present work, sufficient criteria for steerability will be constructed. Some known results in Refs.~\cite {Werner,Wiseman1} are required in the derivation of LSI for the continuous setting in the following. Therefore, for the convenience of readability, we would like to give a brief review of the  the Werner states~\cite{Werner} and isotropic states~\cite{Wiseman1}. Especially, some formulas, which are very useful in this work, will be introduced below.

\subsection{Werner states}

The  Werner states are defined as~\cite{Werner,JWD}
\begin{equation}
W^w_d=\frac{d-1+w}{d-1}\frac{I_d \otimes I_d}{d^2}-\frac{w}{d-1}\frac{\mathbf{V}}{d},
\end{equation}
with $0\leqslant w\leqslant 1$, and $\mathbf{V}$ is the ``flip" operator $\mathbf{V}\vert\psi\rangle\otimes\vert\phi\rangle=\vert\phi\rangle\otimes\vert\psi\rangle$. Werner states are nonseparable iff $w>1/(d+1).$
If Alice performs a projective measurement $\Pi^{A}_a=\vert a\rangle\langle a\vert,\ \forall a\in\{0,1,...,d-1\}$  on her side, the unnormalized  conditional state on Bob's side is
\begin{equation}
\label{constate}
\tilde{\rho}^A_a=\frac{d-1+w}{d(d-1)}\frac{I_d}{d}-\frac{w}{d(d-1)}\vert a\rangle\langle a\vert.
\end{equation}
It was shown that the original derivation by Werner in Ref.~\cite{Werner} can be equivalently expressed in terms of steering~\cite{Wiseman1}. Denote the $d$-dimensional unitary group by $\mathrm{U}(d)$, and with a unitary operator  $\hat{U}_{\omega}\in\mathrm{U}(d)$, an state $\vert\psi_{\omega}\rangle$ can be expressed as $\vert\psi_{\omega}\rangle=\hat{U}_{\omega}\vert 0\rangle$, where $\vert 0\rangle$ is a fixed state in the $d$-dimensional Hilbert space and $\omega$ represents the group parameters. The complete set of pure states in the $d$-dimensional system is denoted by $F^\star\equiv\{\vert\psi_{\omega}\rangle\langle\psi_{\omega}\vert d\mu_{\mathrm{Haar}}(\omega)\}$, with $d\mu_{\mathrm{Haar}}(\omega)$ the Harr measure on the group $\mathrm{U}(d)$. If Alice is trying to simulate the conditional state above, the optimal set of LHS should be $F^\star$ \cite{Wiseman1}. Formally, the simulation can be described as
\begin{equation}
\tilde{\rho}^A_a=\int d\omega\Omega(\omega)\mathfrak{p}(a\vert A, \psi_{\omega})\vert\psi_{\omega}\rangle\langle\psi_{\omega}\vert,
\end{equation}
with the constraint $\sum_{a=0}^{d-1}\mathfrak{p}(a\vert A, \psi_{\omega})=1$ and a probability distribution $\Omega(\omega)$. For an explicit conditional state in Eq.~\eqref{constate}, the optimal choice of $\{\mathfrak{p}(a\vert A, \psi_{\omega})\}$ is
\begin{equation}
\label{optp}
\mathfrak{p}^\star(a\vert A, \psi_{\omega})=\left\{\begin{array}{l}
                                1~~\mathrm{if}~\langle\psi_{\omega}\vert\hat{\Pi}^A_a\vert\psi_\omega\rangle<\langle\psi_{\omega}\vert\hat{\Pi}^A_{a'}\vert\psi_{\omega}\rangle,~a\neq a'\\
                                0~~\mathrm{otherwise}
                              \end{array}\right..
\end{equation}
As shown by Werner \cite{Werner}, for any positive normalized distribution $\mathfrak{p}(a\vert A, \psi_{\omega})$, there should be
\begin{equation}
\label{inequality1}
\langle a\vert\int d\mu_{\mathrm{Harr}}(\omega)\vert \psi_{\omega}\rangle\langle \psi_{\omega}\vert \mathfrak{p}(a\vert A, \psi_{\omega})\vert a\rangle\geqslant\frac{1}{d^3}.
\end{equation}
The equality is attained for the optimal $\mathfrak{p}^\star(a\vert A, \psi_{\omega})$ specified in Eq.~\eqref{optp}. From it, it can be found that Alice cannot simulate the conditional state in Eq.~\eqref{constate} iff $(1-w)/d^2<1/d^3$.

\subsection{Isotropic states}
The isotropic states, which were introduced in Ref.~\cite{Horo}, can be parameterized similarly to the Werner states with a mixing parameter $\eta$,
\begin{equation}
\label{isotropic}
W^{\eta}_d=(1-\eta)\frac{I_d \otimes I_d}{d^2}+\eta \mathbf{P}_+.
\end{equation}
Here $\mathbf{P}_+=\vert\psi_+\rangle\langle\psi_+\vert$, where $\vert\psi_+\rangle=\sum_{i=1}^d\vert i\rangle\otimes\vert i\rangle/\sqrt{d}$  is a maximally entangled state. For $d=2$, the isotropic states are identical to Werner states up to local unitary transformations.
These states are entangled if $\eta>1/(d+1)$.

If Alice makes a projective measurement, the conditional state for Bob is
\begin{equation}
\label{constate2}
\rho^A_a=\frac{1-\eta}{d}\frac{I_d}{d}+\frac{\eta}{d}\vert a\rangle\langle a \vert.
\end{equation}
When Alice tries to simulate this state, the ensemble $F^\star:=\{\vert\psi_{\omega}\rangle\langle\psi_{\omega}\vert d\mu_{\mathrm{Haar}}(\omega)\}$ has been proved to be the optima one~\cite{Wiseman1}. Especially, the choice of the $\mathfrak{p}(a\vert A, \psi_{\omega})$
\begin{equation}
\label{optp2}
\mathfrak{p}^\star(a\vert A, \psi_{\omega})=\left\{\begin{array}{l}
                                1~~\mathrm{if}~\langle\psi_{\omega}\vert\hat{\Pi}^A_a\vert\psi_\omega\rangle>\langle\psi_{\omega}\vert\hat{\Pi}^A_{a'}\vert\psi_{\omega}\rangle,~a\neq a'\\
                                0~~\mathrm{otherwise}
                              \end{array}\right.
\end{equation}
is optimal for Alice to simulate the conditional states in Eq.~\eqref{constate2}. It has been found that for any positive normalized distribution
$\{\mathfrak{p}(a\vert A, \psi_{\omega})\}$,
\begin{equation}
\label{inequality}
\langle a\vert\int d\mu_{\mathrm{Harr}}(\omega)\vert \psi_{\omega}\rangle\langle \psi_{\omega} \vert\mathfrak{p}(a\vert A, \psi_{\omega})\vert a\rangle\leqslant\frac{H_d}{d^2},
\end{equation}
where $H_d=\sum_{n=1}^d(1/n)$ is the Harmonic series and the equality is attained for the optimal $\mathfrak{p}^\star(a\vert A, \psi_{\omega})$ specified in Eq.~\eqref{optp2}. Therefore, Alice cannot simulate the conditional states iff $\eta/d+(1-\eta)/d^2> H_d/d^2$~\cite{Wiseman1}.

\section{Nonsteering threshold}
\label{Sec3}

\subsection{ Sufficient criteria for steering}
The conditional state $\tilde{\rho}^{a}_{\mu}$ on Bob's side can be measured with a set of rank-one projective operators $\{\hat{M}^a_{\mu}\}$,
$\hat{M}^a_{\mu}\equiv\hat{\Phi}^a_{\mu}=\vert \phi^a_{\mu}\rangle\langle\phi^a_{\mu}\vert,\langle\phi^a_{\mu}\vert\phi^b_{\mu}\rangle=\delta_{ab}$, $\sum_{a=0}^{d-1}\hat{M}_{\mu}^{a}=I_{d}$, and the fidelity for the $\mu$-th run of experiment  is defined as
\begin{equation}
F_{\mu}=\sum_{a=0}^{d-1}\mathrm{Tr}\left[\tilde{\rho}^{a}_{\mu}\hat{\Phi}^a_{\mu}\right].
\end{equation}
 Let $\langle A\otimes B  \rangle =\mathrm{Tr}(A\otimes B W)$ be the expectation value of the operator $A\otimes B$, and in experiment,  $F_{\mu}$ can be measured as
\begin{equation}
F_{\mu}=\sum_{a=0}^{d-1}\left\langle \hat{\Pi}^a_{\mu}\otimes \hat{\Phi}^a_{\mu}\right\rangle.
\end{equation}
Assume the probability of the $\mu$-th measurement performed by Alice is $q_{\mu}$, $\sum_{\mu=1}^N q_{\mu}=1$, and the averaged fidelity $\bar{F}$ can be defined
\begin{equation}
\bar{F}\equiv\sum_{\mu=1}^N q_{\mu}F_{\mu}.
\end{equation}

The averaged fidelity plays an important role in the detection of entanglement. Let us recall the decomposition in Eq.~\eqref{decoposition}, and now the channel $\varepsilon$ is restricted to be an EB one, denoted by $\varepsilon_{\mathrm{EB}}$. With the assemblage resulted from the EB channel, $\{\tilde{\rho }^a_{\mu}=\varepsilon_{\mathrm{EB}}(\sqrt{\rho_{\mathrm{A}}}(\hat{\Pi}_{\mu}^a)^*\sqrt{\rho_{\mathrm{A}}})\}$, and following the above definitions, one has the fidelity $F_\mu^{\mathrm{EB}}=\sum_{a=0}^{d-1}\mathrm{Tr}[\hat{\Phi}^a_{\mu}\varepsilon_{\mathrm{EB}}(\sqrt{\rho_{\mathrm{A}}}
(\hat{\Pi}_{\mu}^a)^*\sqrt{\rho_{\mathrm{A}}})]$ and the averaged fidelity $F^{\mathrm{EB}}_{\mathrm{avg}}=\sum_{\mu}q_{\mu}F_\mu^{\mathrm{EB}}$.
The classical fidelity threshold (CFT) can be defined as $\mathfrak{F}_{\mathrm{CFT}}=\max_{\varepsilon_{\mathrm{EB}}}F^{\mathrm{EB}}_{\mathrm{avg}}$, where the maximum is taken over the set of all EB channels $\{ \varepsilon_{\mathrm{EB}}\}$~\cite{Barnett1,Fuchs,Massar,Horo,Adesso1,Chir,Namiki3,Chir1,Namiki4}. If the experiment data $\bar{F}$ exceeds this threshold, $\bar{F}>\mathfrak{F}_{\mathrm{CFT}}$, one may conclude that the channel $\varepsilon$ in Eq.~\eqref{decoposition} cannot be a EB channel and the state $W$ is an entangled state.

The above idea to detect entanglement is heuristic and sheds light on the detection of steering. Similarly, by taking the averaged fidelity as the steering parameter, a steering inequality can also be constructed by just considering the measurement performed by Bob~\cite{can22,sau,Joness}. Assume that the assemblage $\{\tilde{\rho}^a_{\mu}\}$ introduced in Eq.~\eqref{df}  has an LHS decomposition in Eq.~\eqref{tilderho}, and one can have $\mathrm{Tr}[\tilde{\rho}^{a}_{\mu}\hat{\Phi}^a_{\mu}]=\int d\xi\Omega(\xi)\mathfrak{p}(a\vert\mu,\xi)\mathrm{Tr}(\rho_{\xi}\hat{\Phi}^{a}_{\mu})$. With the definition of averaged fidelity above, one may
 introduce
 an averaged fidelity
\begin{equation}
F^{\mathrm{LHS}}_{\mathrm{avg}}\equiv\sum_{\mu=1}^N\sum_{a=0}^{d-1}q_{\mu}
\int d\xi\Omega(\xi)\mathfrak{p}(a\vert\mu,\xi)\mathrm{Tr}(\rho_{\xi}\hat{\Phi}^{a}_{\mu})
\end{equation}
for the case where the assemblage $\{\tilde{\rho}^{a}_{\mu}\}$ admits a LHS model.
Formally, it can be rewritten as  $F^{\mathrm{LHS}}_{\mathrm{avg}}\equiv\int d\xi\Omega(\xi)\mathrm{Tr}(\rho_{\xi}\bar{\rho})$,  with $\bar{\rho}$ defined as
\begin{equation}
\label{rhobar}
\bar{\rho}=\sum_{\mu=1}^N\sum_{a=0}^{d-1}q_{\mu}\mathfrak{p}(a\vert\mu,\xi)\hat{\Phi}^{a}_{\mu},
\end{equation}
with the probability $\mathfrak{p}(a\vert\mu,\xi)$ interpreted as the value of $\hat{\Pi}^a_{\mu}$ in the local hidden variable (LHV) model. $\bar{\rho}$ is a density matrix, and formally, can be expanded as $\bar{\rho }=\sum_{\nu}\lambda_{\nu}\vert \lambda_\nu\rangle\langle \lambda_{\nu}\vert$, with $\lambda_{\nu}$ the eigenvalues and $\vert \lambda_{\nu}\rangle$ the corresponding eigenvectors.
Defining $\lambda^{\max}=\max_{\mathfrak{p}(a\vert\mu,\xi)}\max_{\mu}\{\lambda_{\mu}\}$, and together with the facts $\mathrm{Tr}[\rho_{\xi}\bar{\rho}]\leqslant\lambda^{\max}$ and $\int \Omega(\xi)d\xi=1$, one can conclude that $\lambda^{\max}$ is an upper bound of $F^{\mathrm{LHS}}_{\mathrm{avg}}$, say, $\lambda^{\max}\geqslant F^{\mathrm{LHS}}_{\mathrm{avg}}$.
From the definition of unsteerable states, we know that the assemblage resulted from the unsteerable state always admits an LHS model. Therefore,
$\lambda^{\max}$ also can be interpreted as the upper bound of the averaged fidelity derived from the unsteerable states, when the measurement on Bob's has been fixed as $\{ q_{\mu},M^a_{\mu}\}$. To emphasize this property of  $\lambda^{\max}$, we define it as the nonsteering threshold (NST) and denote it  by the symbol $\mathfrak{F}_{\mathrm{NST}}^+$ hereafter,
\begin{equation}
\label{NST+}
\mathfrak{F}_{\mathrm{NST}}^+(\{q_{\mu},M^a_{\mu}\})=\max_{\vert\phi\rangle}\max_{\mathfrak{p}(a\vert\mu,\xi)}\langle\phi\vert\bar{\rho}\vert\phi\rangle.
\end{equation}

In a similar way, the minimum eigenvalue of $\bar{\rho}$ is the other NST, and denoted by $\mathfrak{F}_{\mathrm{NST}}^-$ hereafter
\begin{equation}
\label{NST-}
\mathfrak{F}_{\mathrm{NST}}^-(\{q_{\mu},M^a_{\mu}\})=\min_{\vert\phi\rangle}\min_{\mathfrak{p}(a\vert\mu,\xi)}\langle\phi\vert\bar{\rho}\vert\phi\rangle.
\end{equation}
With $ \mathrm{Tr}[\rho_{\xi}\bar{\rho}]\geqslant\mathfrak{F}^-_{\mathrm{NST}}$ and $\int \Omega(\xi)d\xi=1$, one can conclude that
$\mathfrak{F}^-_{\mathrm{NST}}$ is a lower bound of $F^{\mathrm{LHS}}_{\mathrm{avg}}$, say, $\mathfrak{F}^-_{\mathrm{NST}}\leqslant F^{\mathrm{LHS}}_{\mathrm{avg}}$. Therefore, an LSI can be defined
\begin{equation}
\label{LSI}
\mathfrak{F}_{\mathrm{NST}}^-(\{q_{\mu},M^a_{\mu}\})\leqslant\bar{F}\leqslant\mathfrak{F}_{\mathrm{NST}}^+(\{q_{\mu},M^a_{\mu}\}).
\end{equation}
 Since the following two conditions---(a) the state is steerable from Alice to Bob and (b) the set of measurements $\{\Pi^a_{\mu}\}$ performed by Alice is incompatible---are necessary so that the assemblage $\{{\tilde{\rho}}^{a}_{\mu}\}$ does not admit an LHS  model, one may conclude that the violation of the steering inequality, is a sufficient condition for Bob to make the statements (a) and (b).

To show a state $W$ is steerable from Alice to Bob, the extremal values of the averaged fidelity should be considered. For a fixed measurement $\{\hat{\Phi}^a_{\mu}\}_{a=0}^{d-1}$ performed by Bob, let $F_{\mu}^+$ ($F_{\mu}^-$) be the maximum (minimum) value of $F_{\mu}$ with the corresponding measurement $\{\hat{\Pi}^a_\mu\}_{a=0}^{d-1}$ performed by Alice. The extremal values of the fidelity are
\begin{equation}
\bar{F}^{\pm}(\{q_{\mu},M^a_{\mu}\})=\sum_{\mu=1}^N q_{\mu}F_{\mu}^{\pm},
\end{equation}
and obviously, $\bar{F}^-\leqslant\bar{F}\leqslant\bar{F}^+$. Now, two types of steering criteria can be introduced. For the reason which will be clarified in the following, one can define the Wiseman-Jones-Doherty (WJD) type criterion
\begin{equation}
\label{WJDtype}
\bar{F}^+(\{q_{\mu},M^a_{\mu}\})> \mathfrak{F}_{\mathrm{NST}}^+(\{q_{\mu},M^a_{\mu}\}),
\end{equation}
and the Werner-type one
\begin{equation}
\label{Wernertype}
\bar{F}^-(\{q_{\mu},M^a_{\mu}\})<\mathfrak{F}_{\mathrm{NST}}^-(\{q_{\mu},M^a_{\mu}\}).
\end{equation}
The two criteria above are independent, which means that if either is verified, the state $W$ is demonstrated to be steerable from Alice to Bob. It will be shown that both types of the criteria should be considered for the high-dimensional system ($d>2$).

\subsection{The optimal eigenvectors}
First, let us consider the probabilistic LHV model. For the $\mu$-th measurement $\{\Pi^{a}_{\mu}\}$, $\sum_{a=0}^{d-1}\Pi^{a}_{\mu}=I_d$, and
\begin{equation}
0\leqslant\mathfrak{p}(a\vert\mu,\xi)\leqslant 1,~\sum_{a=0}^{d-1}\mathfrak{p}(a\vert\mu,\xi)=1.
\end{equation}
From Eqs.~\eqref{rhobar} and~\eqref{NST+}, a quantity $f_{\mu}(\phi)$ can be introduced
\begin{equation}
\label{quantityf}
f_{\mu}(\phi)=\langle\phi\vert\sum_{a=0}^{d-1} \mathfrak{p}(a\vert\mu,\xi)\hat{\Phi}^{a}_{\mu}\vert\phi\rangle,
\end{equation}
as a function of $\vert\phi\rangle$, and for a fixed $\vert\phi\rangle$, its maximum value,
\begin{equation}
f^{\max}_{\mu}(\phi)=\max_a\langle\phi\vert\hat{\Phi}^{a}_{\mu}\vert\phi\rangle,
\end{equation}
can be obtained with the optimal choice of the probabilities $\{\mathfrak{p}(a\vert\mu,\xi)\}$,
\begin{equation}
\label{optpro}
\mathfrak{p}^\star(a\vert\mu,\xi)=\left\{\begin{array}{l}
                                1~~\mathrm{if}~\langle\phi\vert\hat{\Phi}^a_{\mu}\vert\phi\rangle>\langle\phi\vert\hat{\Phi}^{a'}_{\mu}\vert\phi\rangle,~a\neq a'\\
                                0~~ \mathrm{otherwise}
                              \end{array}\right.,
\end{equation}
where $a, a'\in\{0,1, ..., d-1\}$. $\mathfrak{F}^+_{\mathrm{NST}}$ can be rewritten as
\begin{equation}
\mathfrak{F}^+_{\mathrm{NST}}=\max_{\vert\phi\rangle}\sum_{\mu=1}^N q_{\mu}f^{\max}_{\mu}(\phi).
\end{equation}
Next, one may seek the optimal state $\vert\phi_+\rangle$ corresponding to the largest eigenvalue of $\bar{\rho}$, and then, the result can be formally expressed as
\begin{equation}
\label{NST+2}
\mathfrak{F}^+_{\mathrm{NST}}=\sum_{\mu=1}^N\sum_{a=0}^{d-1}q_{\mu}\langle\phi_+\vert\mathfrak{p}^\star(a\vert\mu,\xi)\hat{\Phi}^a_{\mu}\vert\phi_+\rangle.
\end{equation}

On the other hand, $\mathfrak{F}^-_{\mathrm{NST}}$ can be derived similarly. With the optimal probabilities
\begin{equation}
\mathfrak{p}^\star(a\vert\mu,\xi)=\left\{\begin{array}{l}
                                1~~\mathrm{if}~\langle\phi\vert\hat{\Phi}^a_{\mu}\vert\phi\rangle<\langle\phi\vert\hat{\Phi}^{a'}_{\mu}\vert\phi\rangle,~a\neq a'\\
                                0~~\mathrm{otherwise }
                              \end{array}\right.,
\end{equation}
the same quantity $f_{\mu}(\phi)$ in Eq.~\eqref{quantityf} achieves its minimum value
\begin{equation}
f^{\min}_{\mu}(\phi)=\min_a\langle\phi\vert\hat{\Phi}^{a}_{\mu}\vert\phi\rangle.
\end{equation}
Therefore,
\begin{equation}
\mathfrak{F}^-_{\mathrm{NST}}=\min_{\vert\phi\rangle}\sum_{\mu=1}^N q_{\mu}f^{\min}_{\mu}(\phi),
\end{equation}
and by choosing the optimal vector $\vert\phi_-\rangle$ corresponding to the minimum eigenvalue of $\bar{\rho}$, one can come to the final result
\begin{equation}
\label{NST-1}
\mathfrak{F}^-_{\mathrm{NST}}=\sum_{\mu=1}^N\sum_{a=0}^{d-1}q_{\mu}\langle\phi_-\vert\mathfrak{p}^\star(a\vert\mu,\xi)\hat{\Phi}^a_{\mu}\vert\phi_-\rangle.
\end{equation}
Until now, we have considered the case where the number of the experiment settings is finite, and the above conclusions can be easily generalized to the case where the experiment settings are continuous.

\subsection{Deterministic LHV model}
In the above discussion, a general protocol to calculate NSTs through finding the optimal eigenvectors has been constructed. From the optimal choice of $\{\mathfrak{p}(a\vert\mu,\xi)\}$, it is shown that the NSTs are unchanged if a deterministic LHV is applied. For the $\mu$-th measurement $\{\Pi^{a}_{\mu}\}_{a=0}^{d-1}$, $\sum_{a=0}^{d-1}\Pi^{a}_{\mu}=I_d$, and
\begin{equation}
\mathfrak{p}(a\vert\mu,\xi)\in\{0,1\},~~\sum_{a=0}^{d-1}\mathfrak{p}(a\vert\mu,\xi)=1.
\end{equation}
So, one may have another way to derive the NSTs, shown in the following. For the measurements $\{q_{\mu},\hat{\Phi}^a_{\mu}\}$, a density matrix can be introduced
\begin{equation}
\bar{\rho}_{k_1,k_2,...,k_N}=\sum_{\mu=1}^N q_{\mu}\hat{\Phi}_{\mu}^{k_{\mu}},
\end{equation}
where $k_\mu\in\{0,1,...,d-1\}$ for all $\mu=1,2,...,N$. There are totally $d^N$ matrices of such kind. For $\bar{\rho}_{k_1,k_2,...,k_N}$, the largest eigenvalue and the minimum eigenvalue are
\begin{eqnarray}
\lambda^{\max}_{k_1,k_2,...,k_N}&=&\max_{\vert\phi\rangle}\langle\phi\vert \bar{\rho}_{k_1,k_2,...,k_N}\vert\phi\rangle,\\
\lambda^{\min}_{k_1,k_2,...,k_N}&=&\min_{\vert\phi\rangle}\langle\phi\vert \bar{\rho}_{k_1,k_2,...,k_N}\vert\phi\rangle,
\end{eqnarray}
respectively. The NSTs can be expressed as
\begin{eqnarray}
\mathfrak{F}^+_{\mathrm{NST}}&=&\max _{k_1,k_2,...,k_N}\lambda^{\max}_{k_1,k_2,...,k_N},\\
\mathfrak{F}^-_{\mathrm{NST}}&=&\min_{k_1,k_2,...,k_N}\lambda^{\min}_{k_1,k_2,...,k_N}.
\end{eqnarray}

As an illustration, let us consider a two-settings case as a specific example. Two sets of orthogonal basis $\{\vert\phi^a_{1}\rangle\}$ and $\{\vert\phi^b_{2}\rangle\}$ with $a,b=0,1,...,d-1$ can be chosen, which are related by a unitary matrix $U$. $U_{ab}$ are matrix elements and $\vert\phi^b_2\rangle=\sum_{a=0}^{d-1} U_{ba}\vert\phi^a_1\rangle$. Fixing the probability for each setting as $q_1=q_2=1/2$ and with the deterministic LHV model, a series of states $\bar{\rho}_{a,b}=(\hat{\Phi}^{a}_1+\hat{\Phi}^b_2)/2$ can be introduced and NST can be obtained:
\begin{equation}
\mathfrak{F}^+_{\mathrm{NST}}=\max_{a,b}\lambda^{\max}_{a,b},~~\mathfrak{F}^-_{\mathrm{NST}}=\min_{a,b}\lambda^{\min}_{a,b}.
\end{equation}
For a mixed state $\rho=(\vert e_1\rangle\langle e_1\vert+\vert\varphi\rangle\langle \varphi\vert)/2$, where the state $\vert\varphi\rangle=s\vert e_1\rangle+\sqrt{1-\vert s\vert^2} \vert e_2\rangle$ with two orthogonal bases $\vert e_1\rangle$ and $\vert e_2\rangle$, its maximum eigenvalue is $\lambda^{\max}(\rho)=(1+\vert s\vert)/2$. Based on this fact, $\lambda^{\max}_{a,b}=(1+\vert U_{ab}\vert)/2$ and
\begin{equation}
\mathfrak{F}^+_{\mathrm{NST}}=\frac{1}{2}(1+\max_{a,b}\vert U_{ab}\vert).
\end{equation}
One can select out the optimal element $U^{\mathrm{opt}}_{ab}$, whose modulus $\vert U^{\mathrm{opt}}_{ab}\vert$ has the largest value,  from all the unitary matrix elements. Then, $\mathfrak{F}^+_{\mathrm{NST}}=(1+ \vert U^{\mathrm{opt}}_{ab}\vert)/2$. Note that each $\bar{\rho}_{a,b}$ is a density matrix in $d$-dimensional system, and it has a total number of $d$ eigenvalues. From the definition above, $\bar{\rho}_{a,b}$  is composed of two pure states, it has two nonzero eigenvalues, $\frac{1}{2}(1\pm\vert U_{ab}\vert)$,  and a number of $d-2$ eigenvalues to be zero. Therefore, $\lambda^{\min}_{a,b}=\frac{1}{2}(1-\vert U_{ab}\vert)$ for $d=2$, and $\lambda^{\min}_{a,b}=0$ for  $d>2$. With the definition
 $\mathfrak{F}^-_{\mathrm{NST}}=\min_{a,b}\lambda^{\min}_{a,b}$,
  we have
\begin{equation}
\mathfrak{F}^-_{\mathrm{NST}}=\left\{\begin{array}{l}
                                \frac{1}{2}(1-\vert U^{\mathrm{opt}}_{ab}\vert)~~\mathrm{if}~d=2\\
                               0~~~~~~~~~~~~~~~~~~~~~~~\mathrm{if}~d>2
                              \end{array}\right..
\end{equation}

It is known that a set of mutually unbiased bases (MUBs) consists of two or more orthonormal bases $\{\vert\phi_{x}^a\rangle\}$ in a $d$-dimensional Hilbert space satisfying
\begin{equation}
\left\vert\langle\phi^a_x\right\vert\phi^b_y\rangle\vert^2=\frac{1}{d},~\forall a,b\in\{0,1,...,d-1\},~x\neq y,
\end{equation}
for all $x$ and $y$~\cite{mubs}. Formally,  one can introduce a unitary matrix $U$ with $\vert\phi^a_x\rangle=\sum_{a=0}^{d-1}U_{ab}\vert\phi^b_y\rangle$. From the definition for MUBs, $\vert U_{ab}\vert=1/\sqrt{d},\ \forall a, b \in\{0,1, ...,d-1\}$. With $\mathfrak{F}^+_{\mathrm{NST}}=(1+1/\sqrt{d})/2$ and the averaged fidelity $\bar{F}=\frac{1}{2}\sum_{a=0}^{d-1}\sum_{\mu=1}^2\langle\hat\Pi_{\mu}^a\otimes\hat{\Phi}^{a}_{\mu}\rangle$, one can have the WJD-type steering criterion
\begin{equation}
\sum_{a=0}^{d-1}\sum_{\mu=1}^2\langle\hat\Pi_{\mu}^a\otimes\hat{\Phi}^{a}_{\mu}\rangle >1+\frac{1}{\sqrt{d}},
\end{equation}
with $\hat\Phi^a_{\mu}$ one of MUBs. This result has appeared in previous works with different approaches~\cite{Li,Zeng}.

\subsection{Geometric steering inequality}
Here, the geometric averaged fidelity, which is related to the averaged fidelity $\bar{F}$ in a simple way, can be defined as
\begin{equation}
\label{geoavg}
\bar{f}\equiv\frac{d\bar{F}-1}{d-1}.
\end{equation}
Correspondingly, one can define the so-called geometric NSTs,
\begin{equation}
\mathfrak{g}^{\pm}_{\mathrm{NST}}(\{q_{\mu},M^a_{\mu}\})=\frac{d\mathfrak{F}^{\pm}_{\mathrm{NST}}(\{q_{\mu},M^a_{\mu}\})-1}{d-1},
\end{equation}
and the criteria about Eq.~\eqref{LSI} can be equivalently expressed as the following: If the geometric inequality,
\begin{equation}
\mathfrak{g}^{-}_{\mathrm{NST}}\leqslant\bar{f}\leqslant\mathfrak{g}^{+}_{\mathrm{NST}},
\end{equation}
is violated, the state $W$ is steerable from Alice to Bob. This type of inequality is convenient for the qubit case. With $\sigma_x$, $\sigma_y$, and $\sigma_z$ the Pauli matrices and a three-dimensional Bloch vector $\mathbf{r}=r_y\mathbf{\hat{x}}+r_y\mathbf{\hat{y}}+r_z \mathbf{\hat{z}}$ ($\hat{\mathbf{x}}$, $\hat{\mathbf{y}}$, and $\hat{\mathbf{z}}$ are unit vectors along coordinate axes), a density matrix can be expressed as $\rho=(I_2+\mathbf{r}\cdot\bm{\sigma})/2$, with $\mathbf{r}\cdot\bm{\sigma}=r_x\sigma_x+r_y\sigma_y+r_y\sigma_z$. The geometric length of $\mathbf{r}$ is $\vert \mathbf{r}\vert=\sqrt{r_x^2+r_y^2+r_z^2}$. Furthermore, the measurement results of Alice are usually denoted by $a=+,-$. Then, the measurements performed by Alice can be expressed as $\hat{\Pi}^{\pm}_{\mu}=(I_2\pm\mathbf{\hat{r}}_{\mu}\cdot\bm{\sigma})/2$, and the target states can be written as $\hat{\Phi}^{\pm}_{\mu}=(I_2\pm\mathbf{\hat{n}}_{\mu}\cdot\bm{\sigma})/2$, where $\mathbf{\hat{r}}_{\mu}$ and $\mathbf{\hat{n}}_{\mu}$ are unit vectors. Now, one can define a quantity $\mathfrak{A }(\mu,\xi)=\mathfrak{p}(+\vert\mu,\xi)-\mathfrak{p}(-\vert\mu,\xi)$, and by the constraint $\mathfrak{p}(+\vert\mu,\xi)+\mathfrak{p}(-\vert\mu,\xi)=1$, it can be obtained that $-1\leqslant\mathfrak{A }(\mu,\xi)\leqslant1$. In fact, $\mathfrak{A }(\mu,\xi)$ may be viewed as the predetermined value of the operator $\mathbf{\hat{ r}}_{\mu}\cdot\bm{\sigma}$ in an LHV  model. With the vector $\mathbf{\bar{r}}=\sum_{\mu=1}^N q_{\mu} \mathfrak{A}(a\vert\mu,\xi)\mathbf{\hat{n}}_{\mu}$, the state $\bar{\rho}$ in Eq.~\eqref{rhobar} can be expressed as $\bar{\rho}=(I_2+\mathbf{\bar{\mathbf{r}}}\cdot\bm{\sigma})/2$, and introducing the optimal length of $\mathbf{\bar{r }}$,
\begin{equation}
\vert\mathbf{\bar{r }}\vert_{\mathrm{opt}}=\max_{-1\leqslant\mathfrak{A }(\mu,\xi)\leqslant1}\vert \bar{\mathbf{r}}\vert,
\end{equation}
the geometric NSTs can be obtained as follows
 \begin{equation}
 \mathfrak{g}^{\pm}_{\mathrm{NST}}=\pm\vert\mathbf{\bar{r }}\vert_{\mathrm{opt}}.
 \end{equation}
With Eq.~\eqref{geoavg}, the geometric averaged fidelity can be expressed as $\bar{f}=\sum_{\mu=1}^Nq_{\mu}\langle \mathbf{\hat{r}}_{\mu}\cdot\bm{\sigma}\otimes \mathbf{\hat{n}}_{\mu}\cdot\bm{\sigma}\rangle$, and the geometric steering inequality for the qubit case becomes
\begin{equation}
\label{geoLSI}
-\vert\mathbf{\bar{r }}\vert_{\mathrm{opt}}\leqslant\sum_{\mu=1}^Nq_{\mu}\langle\mathbf{\hat{r}}_{\mu}\cdot\bm{\sigma}\otimes\mathbf{\hat{n}}_{\mu}\cdot\bm{\sigma}\rangle\leqslant\vert\mathbf{\bar{r }}\vert_{\mathrm{opt}}.
\end{equation}

For the deterministic LHV model, $\mathfrak{A}(\mu,\xi)\in\{-1,+1\}$, and introducing $2^N$ vectors $\mathbf{\bar{r}}_{\pm\pm...\pm}= \sum_{\mu=1}^N(\pm q_\mu\mathbf{\hat{n}}_\mu)$, the optimal length of $\mathbf{\bar{r }}$ can be expressed as
\begin{equation}
\vert\mathbf{\bar{r }}\vert_{\mathrm{opt}}=\max_{\pm\pm...\pm}\vert \mathbf{\bar{r}}_{\pm\pm...\pm}\vert.
\end{equation}
The known result in Ref.~\cite{sau} is recovered here.

As a simple example, let us consider the case where Bob's measurements are MUBs: $\mathbf{\hat{n}}\cdot\bm{\sigma}$ and $\mathbf{\hat{n}}_{\bot}\cdot\bm{\sigma}$, where $\mathbf{\hat{n}}$ and $\mathbf{\hat{n}}_{\bot}$ are two orthogonal unit vectors.  With $q(\mathbf{\hat{n}})$
and $q(\mathbf{\hat{n}}_{\bot})$ the probabilities for each measurement, respectively, all the possible four vectors are $\mathbf{\bar{r}}_{\pm\pm}=\pm q(\mathbf{\hat{n}})\mathbf{\hat{n}}\pm q(\mathbf{\hat{n}}_{\bot})\mathbf{\hat{n}}_{\bot}$, and it is easy to calculate the geometric length for each vector $\vert \mathbf{\bar{r}}_{\pm\pm}\vert=\sqrt{ q^2(\mathbf{\hat{n}})+q^2(\mathbf{\hat{n}}_{\bot})}$. Thus, the optimal length is
$\vert\mathbf{\bar{r }}\vert_{\mathrm{opt}}=\sqrt{ q^2(\mathbf{\hat{n}})+q^2(\mathbf{\hat{n}}_{\bot})}$ and the geometric steering inequality above has a more explicit form
\begin{equation}
\label{explicit}
-1\leqslant\frac{q(\mathbf{\hat{n}})\langle \hat{\mathbf{a}}\otimes \mathbf{\hat{n}}\rangle}{\sqrt{q^2(\mathbf{\hat{n}})+q^2(\mathbf{\hat{n}}_{\bot})}}
+\frac{q(\mathbf{\hat{n}}_{\bot})\langle \hat{\mathbf{b}}\otimes \mathbf{\hat{n}}_{\bot}\rangle}{\sqrt{q^2(\mathbf{\hat{n}})+q^2(\mathbf{\hat{n}}_{\bot})}}\leqslant 1,
\end{equation}
where $\langle \hat{\mathbf{a}}\otimes \mathbf{\hat{n}}\rangle=\langle\hat{\mathbf{a}}\cdot\bm{\sigma}\otimes\mathbf{\hat{n}}\cdot\bm{\sigma}\rangle$.

If the CHSH inequality~\cite{chsh},
\begin{equation}
\label{CHSH}
-2\leqslant\langle \mathbf{\hat{a}}\otimes (\mathbf{\hat{n}}_1-\mathbf{\hat{n}}_2)\rangle +\langle \mathbf{\hat{b}}\otimes (\mathbf{\hat{n}}_1+\mathbf{\hat{n}}_2)\rangle\leqslant2
\end{equation}
is violated, the state is Bell-nonlocal. By some algebra shown in Refs.~\cite{pop,h3,Wu}, the CHSH inequality in Eq.~\eqref{CHSH} can take an equivalent form
\begin{equation}
\label{CHSH2}
-1\leqslant\vert\cos\theta\vert\langle \mathbf{\hat{a}}\otimes \mathbf{\hat{n}}\rangle  +\vert\sin\theta\vert \langle \hat{\mathbf{b}} \otimes \mathbf{\hat{n}}_{\bot}\rangle\leqslant1,
\end{equation}
where
\begin{equation}
\mathbf{\hat{n}}_1-\mathbf{\hat{n}}_2=2\vert\cos\theta\vert\mathbf{\hat{n}},~~\mathbf{\hat{n}}_1+\mathbf{\hat{n}}_2=2\vert\sin\theta\vert\mathbf{\hat{n}}_\bot
\end{equation}
It could be found that Eq.~\eqref{CHSH2} is very similar to the criteria in Eq.~\eqref{explicit}. In fact, the two operators
\begin{eqnarray}
\hat{T}_{\mathrm{\mathrm{steer}}}&=&\frac{q(\mathbf{\hat{n}})\hat{\mathbf{a}}\cdot\bm{\sigma}\otimes\mathbf{\hat{n}}\cdot\bm{\sigma}
+q(\mathbf{\hat{n}}_{\bot})\hat{\mathbf{b}}\cdot\bm{\sigma}\otimes\mathbf{\hat{n}}_{\bot}\cdot\bm{\sigma}}
 {\sqrt{q^2(\mathbf{\hat{n}})+q^2(\mathbf{\hat{n}}_{\bot})}},\nonumber\\
 \hat{T}_{\mathrm{CHSH}}&=&\vert \cos\theta\vert\mathbf{\hat{a}}\cdot\bm{\sigma}\otimes\mathbf{\hat{n}}\cdot\bm{\sigma}+\vert \sin\theta\vert \mathbf{\hat{b}}\cdot\bm{\sigma}\otimes\mathbf{\hat{n}}_{\bot}\cdot\bm{\sigma},\nonumber
\end{eqnarray}
are equal $\hat{T}_{\mathrm{\mathrm{steer}}}=\hat{T}_{\mathrm{CHSH}}$, under the following one-to-one mapping
\begin{equation}
\vert \cos\theta\vert=\frac{q(\mathbf{\hat{n}})}
{\sqrt{q^2(\mathbf{\hat{n}})+q^2(\mathbf{\hat{n}}_{\bot})}},\vert \sin\theta\vert=\frac{q(\mathbf{\hat{n}}_{\bot})}
{\sqrt{q^2(\mathbf{\hat{n}})+q^2(\mathbf{\hat{n}}_{\bot})}}.\nonumber
\end{equation}
Based on the results above, one may conclude that if the geometric inequality in Eq.~\eqref{explicit} is violated, the state must be Bell- nonlocal. A similar result has also been found in  \cite{Can3, Girdhar}.

\section{continuous settings}
\label{Sec4}

\subsection{Qubit case}
In the above sections, we have developed a general scheme for constructing LSIs for the discrete case. In this section, two explicit LSIs will be constructed for the case where the measurement performed by Bob has a continuous form. Before the LSIs for an arbitrary dimensional system can be derived, a detailed discussion about the qubit case is required first, and this is useful to show what are necessary to construct the LSIs.

Now, instead of the symbol $\mu$, a three-dimensional unit vector $\hat{\mathbf{n}}=(\sin\theta\cos\phi,\sin\theta\sin\phi,\cos\theta)$ with $0\leqslant\theta\leqslant\pi$ and $0\leqslant\phi<2\pi$, is employed to label Bob's measurement as $\hat{\Phi}^a_{\hat{\mathbf{n}}}$ with $a=+,-$ the outcomes. One can introduce the measure $\frac{1}{4\pi}d^2\hat{\mathbf{n}}\equiv\frac{1}{4\pi}\sin\theta d\theta d\phi$, and certainly, $\frac{1}{4\pi}\int\int d^2\hat{\mathbf{n}}\equiv\frac{1}{4\pi}\int^{2\pi}_0\int^{\pi}_0\sin\theta d\theta d\phi=1$. In general, the measurements $\hat{\Phi}^a_{\hat{\mathbf{n}}}$ have a probability distribution $q(\hat{\mathbf{n}})$, and in this work, we just consider the case that the experimental settings are equal-weighted, say, $q(\hat{\mathbf{n}})=1$. Now, the density matrix in Eq.~\eqref{rhobar} becomes
\begin{equation}
\bar{\rho}=\frac{1}{4\pi}\int\int d^2\hat{\mathbf{n}}\sum_{a}p(a\vert\hat{\mathbf{n}},\xi)\hat{\Phi}^a_{\hat{\mathbf{n}}}.
\end{equation}
Correspondingly, the expressions for its maximum and minimum eigenvalue can be obtained from Eq.~\eqref{NST+} and Eq.~\eqref{NST-}, respectively.

The set of measurements performed by Bob can be denoted by $\{\hat{\Phi}^a_{\hat{\mathbf{n}}},\frac{1}{4\pi} d^2\hat{\mathbf{n}}\}$. This set of measurements has a special property: The optimal vector $\vert\phi_+\rangle$ should be the eigenvector of the measurement which belongs to the set $\{\frac{1}{4\pi}d^2\hat{\mathbf{n}}, \hat{\Phi}^a_{\hat{\mathbf{n}}}\}$. Without loss of generality, one may fix it as an eigenvector of $\hat{\sigma}_z$, $\vert\phi_+\rangle\equiv\vert +\rangle$, where $\hat{\sigma}_z\vert \pm\rangle=\pm\vert\pm\rangle.$
As a comparison, one may recall the case where Bob's measurements are MUBs: $\mathbf{\hat{n}}\cdot\mathbf{{\bm{\sigma}}}$ and $\mathbf{\hat{n}}_{\bot}\cdot\mathbf{{\bm{\sigma}}}$, where $\vert\phi_+\rangle$ should be the eigenvector of $\bar{\mathbf{r}}_{\pm\pm}\cdot\mathbf{{\bm{\sigma}}}$. However, this property does not hold anymore. With $\hat{U}_{\hat{\mathbf{n}}}$ a unitary matrix transforming $\vert +\rangle$ to a state represented by a unit Bloch vector $\hat{\mathbf{n}}$, $\vert \phi_{\hat{\mathbf{n}}}\rangle=\hat{U}_{\hat{\mathbf{n}}}\vert +\rangle$, one may rewrite $\hat{\Phi}^a_{\hat{\mathbf{n}}}=\hat{U}_{\hat{\mathbf{n}}}^{\dagger}\vert a\rangle\langle a\vert \hat{U}_{\hat{\mathbf{n}}}$, and obtain a complete set of pure states $\{\vert \phi_{\hat{\mathbf{n}}}\rangle\}$. By some simply algebra, $\langle +\vert\hat{\Phi}^a_{\hat{\mathbf{n}}}\vert +\rangle=\langle a\vert \phi_{\hat{\mathbf{n}}}\rangle\langle
\phi_{\hat{\mathbf{n}}}\vert a\rangle$, and Eq.~\eqref{NST+2} becomes
\begin{eqnarray}
\label{NST+3}
\mathfrak{F}^+_{\mathrm{NST}}&=&\frac{1}{4\pi}[\langle +\vert(\int\int d^2\hat{\mathbf{n}}\mathfrak{p}^\star(+\vert \hat{\mathbf{n}},\xi)\vert \phi_{\hat{\mathbf{n}}}\rangle\langle\phi_{\hat{\mathbf{n}}}\vert) \vert +\rangle \nonumber\\
&&+\langle -\vert(\int\int d^2\hat{\mathbf{n}}\mathfrak{p}^\star(-\vert \hat{\mathbf{n}},\xi)\vert \phi_{\hat{\mathbf{n}}}\rangle\langle
\phi_{\hat{\mathbf{n}}}\vert) \vert -\rangle],
\end{eqnarray}
with the optimal probabilities
\begin{equation}
\label{opt1}
\mathfrak{p}^\star(a,\vert\hat{\mathbf{n}},\xi)=\left\{\begin{array}{l}
                                1~~\mathrm{if}~\langle a\vert\hat{\Phi}_{\hat{\mathbf{n}}}\vert a\rangle>\langle a'\vert\hat{\Phi}_{\hat{\mathbf{n}}}\vert a'\rangle,~a\neq a'\\
                                0~~\mathrm{otherwise}
                              \end{array}\right.,
\end{equation}
where $a, a'\in\{+,-\}$, and $\hat{\Phi}_{\hat{\mathbf{n}}}=\vert\phi_{\hat{\mathbf{n}}}\rangle\langle \phi_{\hat{\mathbf{n}}}\vert$. Now, only the pure states on the northern hemisphere of the Bloch sphere ($0\leqslant\theta<\pi/2$) contribute to the first term in Eq.~\eqref{NST+3},
\begin{eqnarray}
\frac{1}{4\pi}[\langle &+&\vert(\int\int d^2\hat{\mathbf{n}}\mathfrak{p}^\star(+\vert \hat{\mathbf{n}},\xi)\vert \phi_{\hat{\mathbf{n}}}\rangle\langle\phi_{\hat{\mathbf{n}}}\vert) \vert +\rangle]\nonumber\\
&=&\frac{1}{2}\int_{0}^{\pi/2}\sin\theta d\theta\frac{1}{2}(1+\cos\theta)=\frac{3}{8},
\end{eqnarray}
while, only the pure states on the southern hemisphere of the Bloch sphere contribute to the second term in Eq.~\eqref{NST+3},
\begin{eqnarray}
\frac{1}{4\pi}[\langle &-&\vert(\int\int d^2\hat{\mathbf{n}}\mathfrak{p}^\star(-\vert \hat{\mathbf{n}},\xi)\vert \phi_{\hat{\mathbf{n}}}\rangle\langle\phi_{\hat{\mathbf{n}}}\vert) \vert -\rangle]\nonumber\\
&=&\frac{1}{2}\int_{\pi/2}^{\pi}\sin\theta d\theta\frac{1}{2}(1-\cos\theta)=\frac{3}{8}.
\end{eqnarray}
Collecting the results above together, one can obtain the NST
$\mathfrak{F}^+_{\mathrm{NST}}=3/4$.

With a suitable basis, the optimal vector $\vert\phi_-\rangle$ can be fixed as the eigenvector of $\hat{\sigma}_z$,
$\vert\phi_-\rangle\equiv\vert +\rangle, \hat{\sigma}_z\vert \pm\rangle=\pm\vert\pm\rangle$. In a similar way to the one for deriving $\mathfrak{F}^+_{{\mathrm{NST}}}$, and with the optimal probabilities,
\begin{equation}
\label{opt2}
\mathfrak{p}^\star(a,\vert\hat{\mathbf{n}},\xi)=\left\{\begin{array}{l}
                                1~~\mathrm{if}~\langle a\vert\hat{\Phi}_{\hat{\mathbf{n}}}\vert a\rangle<\langle a'\vert\hat{\Phi}_{\hat{\mathbf{n}}}\vert a'\rangle,~a\neq a'\\
                                0~~\mathrm{otherwise}
                              \end{array}\right.
\end{equation}
where $a,a'\in\{+,-\}$, and one can have $\mathfrak{F}^-_{\mathrm{NST}}=1/4$.
From the derivation above, one can see that the optimal probabilities in Eqs.~\eqref{opt1} and~\eqref{opt2} play an important role in deducing the NSTs. Finally, one can come to a state-independent LSI for the qubit case,
\begin{equation}
\label{LSI1}
\frac{1}{4}\leqslant\bar{F}(\{\frac{1}{4\pi}d^2\hat{\mathbf{n}}, \hat{\Phi}^a_{\hat{\mathbf{n}}}\})\leqslant\frac{3}{4},
\end{equation}
where the measurements by Bob are fixed as $\{\frac{1}{4\pi} d^2\hat{\mathbf{n}}, \hat{\Phi}^a_{\hat{\mathbf{n}}}\}$.

\subsection{High-dimensional case}
With a set of basis vectors $\{\vert a \rangle,a=0,...,d-1\}$, the parameter $\omega$ can be used to label the experiment settings by Bob's measurements, $\hat{\Phi}^a_{\omega}=\hat{U}^{\dagger}_{\omega}\vert a\rangle\langle a \vert \hat{U}_{\omega}$, where $U_\omega$ can take all the unitary operators in the $d$-dimensional unitary group $\mathrm{U}(d)$, and $a$ represents the outcomes. It is assumed that the probability for each measurement is equal-weighted, and a Harr measure $d\mu_{\mathrm{Haar}}(\omega)$ on $\mathrm{U}(d)$ can be introduced,
$\int d\mu_{\mathrm{Haar}}(\omega)\sum_{a=0}^{d-1}\hat{\Phi}^a_{\omega}=I_d$. Formally, the measurements by Bob are denoted by $\{d\mu_{\mathrm{Haar}}(\omega),\hat{\Phi}^a_{\omega}\}$. Meanwhile, $\vert\phi_{\omega}\rangle=\hat{U}_{\omega}\vert 0\rangle$ is a pure state in the $d$-dimensional Hilbert space. Analogously as the qubit case, without loss of generality, the optimal eigenvector is chosen as $\vert\phi_+\rangle\equiv\vert 0\rangle$. Now, Eq.~\eqref{NST+2} may be rewritten into a form more appropriate for the continuous setting
\begin{equation}
\label{NST+4}
\mathfrak{F}^{+}_{\mathrm{NST}}=\sum_{a=0}^{d-1}\langle a\vert(\int d\mu_{\mathrm{Haar}}(\omega)\mathfrak{p}^\star(a\vert \omega,\xi) \vert \phi_{\omega}\rangle\langle \phi_{\omega}\vert)\vert a\rangle,
\end{equation}
where $\langle 0 \vert\hat{\Phi}^a_{\omega}\vert 0\rangle= \langle a  \vert\phi_{\omega}\rangle\langle\phi_{\omega}\vert
a \rangle$ has been applied, and as a generalization of Eq.~\eqref{optpro}, the optimal probabilities are
\begin{equation}
\label{optp3}
\mathfrak{p}^\star(a\vert \omega,\xi)=\left\{\begin{array}{l}
                                1~~\mathrm{if}~\langle \phi_{\omega}\vert a\rangle\langle a\vert\phi_\omega\rangle >\langle \phi_{\omega}\vert a'\rangle\langle a'\vert\phi_\omega\rangle,~a\neq a'\\
                                0~~\mathrm{otherwise}
                              \end{array}\right.
\end{equation}
with $a, a'\in\{0, 1, ...,d-1\}$.

Now, let us come back to the general results about the isotropic states in Sec.~\ref{Sec2}. One may easily verify that the result in Eq.~\eqref{optp3} is similar to the one in Eq.~\eqref{optp2}. As a direct application of the inequality in Eq.~\eqref{inequality}, the NST can be derived as
\begin{equation}
\mathfrak{F}^{+}_{\mathrm{NST}}(\{d\mu_{\mathrm{Haar}},\hat{\Phi}^a_{\omega}\})=\frac{H_d}{d}
\end{equation}
from Eq.~\eqref{NST+4}.

To drive the other NST, similarly, one can fix the optimal eigenvector $\vert\phi_-\rangle\equiv\vert 0\rangle$ and rewrite Eq.~\eqref{NST-1} as
\begin{equation}
\mathfrak{F}^{-}_{\mathrm{NST}}=\sum_{a=0}^{d-1}\langle a\vert(d\mu_{\mathrm{Haar}}(\omega)\mathfrak{p}^\star(a\vert \omega,\xi) \vert \phi_{\omega}\rangle\langle \phi_{\omega}\vert)\vert a\rangle,
\end{equation}
with the optimal probabilities
\begin{equation}
\mathfrak{p}^\star(a\vert \omega,\xi)=\left\{\begin{array}{l}
                                1~~\mathrm{if}~\langle \phi_{\omega}\vert a\rangle\langle a\vert\phi_\omega\rangle <\langle \phi_{\omega}\vert a'\rangle\langle a'\vert\phi_\omega\rangle,~a\neq a'\\
                                0~~\mathrm{otherwise}
                              \end{array}\right.
\end{equation}
where $a, a'\in\{0, 1, ...,d-1\}$. With the inequality in Eq.~\eqref{inequality1}, the other NST can be obtained:
\begin{equation}
\mathfrak{F}^{-}_{\mathrm{NST}}(\{d\mu_{\mathrm{Haar}},\hat{\Phi}^a_{\omega}\})=\frac{1}{d^2}.
\end{equation}
Collecting the above results together, an LSI for the continuous settings $\{d\mu_{\mathrm{Haar}},\hat{\Phi}^a_{\omega}\}$ takes the form
\begin{equation}
\frac{1}{d^2}\leqslant\bar{F}(\{d\mu_{\mathrm{Haar}},\hat{\Phi}^a_{\omega}\})\leqslant\frac{H_d}{d}.
\end{equation}
For any state $W$, if the LSI is violated, the state is verified to be steerable from Alice to Bob, and the measurement performed by Alice is also incompatible. As a special case, the LSI in Eq.~\eqref{LSI1} can be recovered from the general one above with $d=2$.

\section{applications}
\label{Sec5}

\subsection{T-state problem}
An arbitrary two-qubit state can be expressed in the standard form
\begin{equation}
W=\frac{1}{4}(I_2\otimes I_2+\mathbf{a}\cdot\bm{\sigma}\otimes I_2+I_2\otimes\mathbf{b}\cdot\bm{\sigma}+\sum_{jk}T_{jk}\mathbf{\sigma}_i\otimes \mathbf{\sigma}_j),
\end{equation}
where $\mathbf{a}$ and $\mathbf{b}$ are the Bloch vectors for Alice and Bob's reduced states, respectively, and $T$ is the correlation matrix. The T-state is a special class of two-qubit states,
\begin{equation}
W=\frac{1}{4}(I_2\otimes I_2+\sum_{j}t_j\mathbf{\sigma}_j\otimes \mathbf{\sigma}_j),
\end{equation}
where $\mathbf{a}=\mathbf{b}=\bm 0$ and $T$ is a diagonal matrix with $t_j$ the diagonal elements. In 2015, Jevtic~\emph{et. al.} gave a necessary condition of EPR steerability for $T$-states~\cite{jev},
\begin{equation}
\label{Tnec}
\frac{1}{2\pi}\int\int d^2 \hat{\mathbf{n}}\sqrt{\mathbf{\hat{n}}^{\mathrm{T}}T^{2}\hat{\mathbf{n}}}=1.
\end{equation}
The authors also conjectured that the derived condition was precisely the border between steerable and nonsteerable states, and this was later shown analytically~\cite{ngu}. Here, we shall revisit this problem from the view of LSIs.

When Bob's measurement is fixed as $\mathbf{\hat{n}}\cdot\bm{\sigma}$, the expectation $\langle \mathbf{\hat{a}}\otimes \mathbf{\hat{n}}\rangle_+\equiv\max_{\mathbf{\hat{a}}}\langle \mathbf{\hat{a}}\otimes \mathbf{\hat{n}}\rangle$ is the maximum one. Further assume that Bob's measurement is the continuous set $\{\frac{1}{4\pi} d^2\hat{\mathbf{n}}, \hat{\Phi}^a_{\hat{\mathbf{n}}}\}$, and from the definition of the geometric fidelity in Eq.~\eqref{geoavg}, one can have the maximum value of the geometric fidelity $\bar{f}^+\equiv\frac{1}{4\pi}\int\int d^2\hat{\mathbf{n}}\mathbf{\langle\hat{a}}\otimes \mathbf{\hat{n}}\rangle_+$. With the geometric NST $\mathfrak{g}^+_{\mathrm{NST}}=1/2$, which can be directly calculated from Eq.~\eqref{LSI1}, a WJD-type criterion now is constructed
\begin{equation}
\label{g2bit}
\frac{1}{2\pi}\int\int d^2\hat{\mathbf{n}}\mathbf{\langle\hat{a}}\otimes \mathbf{\hat{n}}\rangle_+ > 1.
\end{equation}
This criterion is suitable for any two-qubit state. For the $T$-state, the correlation $\langle \mathbf{\hat{a}}\otimes \mathbf{\hat{n}}\rangle=\mathbf{\hat{a}}\cdot \mathbf{\tilde{n}}$ is the inner product between the two vectors $\mathbf{\hat{a}}=(a_x, a_y, a_z)$ and
$\mathbf{\tilde{n}}=(t_x n_x, t_y n_y, t_z n_z)$.  Via the Cauchy-Schwarz inequality, the optimal choice of $\mathbf{\hat{a}}$ could be $a_i=t_in_i/\sqrt{\sum_i t_i^2n_i^2}$ with $i=x,y,z$. Thus, $\langle \mathbf{\hat{a}}\otimes \mathbf{\hat{n}}\rangle_+=\sqrt{\sum_i t_i^2n_i^2}$, and obviously, $\langle \mathbf{\hat{a}}\otimes \mathbf{\hat{n}}\rangle_+=\sqrt{\mathbf{\hat{n}}^{\mathrm{T}}T^{2}\hat{\mathbf{n}}}$. Therefore, a sufficient condition for the $T$-state to be steerable from Alice to Bob becomes
\begin{equation}
\frac{1}{2\pi}\int \int d^2 \hat{\mathbf{n}}\sqrt{\mathbf{\hat{n}}^{\mathrm{T}}T^{2}\hat{\mathbf{n}}}>1,
\end{equation}
with the equality in Eq.~\eqref{Tnec} the border of it.

The T-state contains only three parameters $t_j$ $(j=1,2,3)$. Naturally, one may ask whether it is possible to obtain an analytical function $g(t_j)=\frac{1}{2\pi}\int \int d^2 \hat{\mathbf{n}}\sqrt{\mathbf{\hat{n}}^{\mathrm{T}}T^{2}\hat{\mathbf{n}}}$. This question has already been discussed in Ref.~\cite{jev}, where Eq.~\eqref{Tnec} has an equivalent form
\begin{equation}
2\pi N_{T}\vert \det T\vert =1,\nonumber
\end{equation}
with $N_{T}$ a surface integral~\cite{jev}. For the special case $t_1=t_2$, an analytical expression for  $N_{T}$ has been found. However, for the general case, it is highly unlikely that one can obtain the desired analytical expression for $g(t_j)$. For the general two-qubit state, it contains more parameters than the T-states, and therefore, when the criterion in Eq.~\eqref{g2bit} is applied, some additional numerical techniques are required to calculate $\bar{f}^+$.

\subsection{Bounds of the general NSTs}
When the state is the isotropic state and a set of measurements $\{q_{\mu}, \hat{\Phi}^a_{\mu}\}$  is used by Bob to detect steering, there is a sufficient criterion that the isotropic state is steerable from Alice to Bob
\begin{equation}
\label{sufcon}
\bar{F}^+_{\eta}( \{q_{\mu}, \hat{\Phi}^a_{\mu}\})> \mathfrak{F}^+_{\mathrm{NST}}( \{q_{\mu}, \hat{\Phi}^a_{\mu}\}),
\end{equation}
where the subscript $\eta$ indicates that the isotropic states are considered. For the $\mu$-th setting of measurements by Bob $\{\hat{\Phi}_{\mu}^a\}_{a=0}^{d-1}$, the conditional states defined in Eq.~\eqref{constate2} can be expressed as
\begin{equation}
\tilde{\rho}^{a}_{\mu}=\frac{1-\eta}{d}\frac{I_d}{d}+\frac{\eta}{d} \hat \Psi^a_{\mu},~a\in\{0,1,...,d-1\}.
\end{equation}
The extreme values of $f_{\eta}\equiv\sum_{a=0}^{d-1} \mathrm{Tr} [\hat{\Phi}^a_{\mu}\tilde{\rho}^{a}_{\mu}]$ will be derived in the following. Obviously, the maximum value $f^{\max}_{\eta}$ can be attained if $\hat{\Phi}^a_{\mu}=\hat{\Psi}^a_{\mu}$, and $f^{\max}_{\eta}=\left[1+(d-1)\eta\right]/d$. The minimum value $f^{\min}_{\eta}$ can be attained by setting $\mathrm{Tr}(\hat{\Phi}^a_{\mu}\hat{\Psi}^{a}_{\mu})=0$, and $f^{\min}_{\eta}=(1-\eta)/d$. Moreover, these extremal values do not depend on the actual form of the measurements $\hat{\Phi}_{\mu}$, and therefore,
\begin{eqnarray}
\bar{F}^+_{\eta}(\{q_{\mu}, \hat{\Phi}^a_{\mu}\})   &=& \frac{1+(d-1)\eta}{d},\nonumber\\
\bar{F} ^-_{\eta}(\{q_{\mu}, \hat{\Phi}^a_{\mu}\})    &=& \frac{1-\eta}{d},\\
\bar{F}^{\pm}_{\eta} (\{q_{\mu}, \hat{\Phi}^a_{\mu}\})    &=& \bar{F}^{\pm}_{\eta} (\{d\mu_{\mathrm{Haar}},\hat{\Phi}^a_{\omega}\}).\nonumber
\end{eqnarray}
However, for the continuous-settings case, the criterion
\begin{equation}
\label{continuous}
\bar{F}^+_{\eta} (\{d\mu_{\mathrm{Haar}},\hat{\Phi}^a_{\omega}\})>\frac{H_d}{d}
\end{equation}
is different from the one in Eq.~\eqref{sufcon}. The WJD threshold $H_d/d$ has been proven to be a tight bound: If it is achieved, the conditional states should admit a LHS model~\cite{Wiseman1}. In other words, the equivalent form of Eq.~\eqref{continuous}
\begin{equation}
\frac{1+(d-1)\eta}{d}>\frac{H_d}{d},
\end{equation}
is a necessary and sufficient condition for the isotropic state to be steerable, while $[1+(d-1)\eta]/d>\mathfrak{F}^+_{\mathrm{NST}}(\{q_{\mu}, \hat{\Phi}^a_{\mu}\})$ is just a sufficient one. Therefore, a state-independent relation does exist
\begin{equation}
\label{NST+5}
\mathfrak{F}^+_{\mathrm{NST}}( \{q_{\mu}, \hat{\Phi}^a_{\mu}\})\geqslant\frac{H_d}{d},
\end{equation}
 where the WJD threshold is a lower bound of the general  $\mathfrak{F}^+_{\mathrm{NST}}( \{q_{\mu}, \hat{\Phi}^a_{\mu}\})$.  This is the reason why we call the criterion in Eq.~\eqref{WJDtype} the WJD-type one.

When the state is the Werner state and the same measurements $\{q_{\mu}, \hat{\Phi}^a_{\mu}\}$ are performed  by Bob, there is a criterion which is sufficient for the Werner state to be steerable from Alice to Bob
\begin{equation}
\bar{F}^-_w( \{q_{\mu}, \hat{\Phi}^a_{\mu}\})< \mathfrak{F}^-_{\mathrm{NST}}( \{q_{\mu}, \hat{\Phi}^a_{\mu}\}),
\end{equation}
where the subscript $w$ is used to indicate that only the Werner state is   considered.
For the $\mu$-th run of experiment, the conditional states,   as they are defined in Eq.~\eqref{constate}, can be expressed as
\begin{equation}
\tilde{\rho}^a_{\mu}=\frac{d-1+w}{d(d-1)}\frac{I_d}{d}-\frac{w}{d(d-1)}\hat{\Psi}^a_{\mu}.
\end{equation}
The extreme values of $f_{w}\equiv\sum_{a=0}^{d-1} \mathrm{Tr} [\hat{\Phi}^a_{\mu}\tilde{\rho}^{a}_{\mu}]$ can be derived as follows. The minimum value $f^{\min}_w$ can be attained if $\hat{\Phi}^a_{\mu}=\hat{\Psi}^a_{\mu}$,  $f^{\min}_w=(1-w)/d$. The maximum value $f^{\max}_{w}$ is obtained by setting $\mathrm{Tr} [\hat{\Phi}^a_{\mu}\hat{\Psi}^{a}_{\mu}]=0$, and $f^{\max}_{w}=(d-1+w)/[d(d-1)]$.  Furthermore, these extremal values do not depend on the actual form of $\hat{\Phi}_{\mu}^a$, and thus, there should be
\begin{eqnarray}
\label{avgf}
\bar{F}^+_{w}(\{q_{\mu}, \hat{\Phi}^a_{\mu}\})   &=& \frac{d-1+w}{d(d-1)},\nonumber\\
\bar{F} ^-_{w}(\{q_{\mu}, \hat{\Phi}^a_{\mu}\})    &=& \frac{1-w}{d},\\
\bar{F}^{\pm}_{w} (\{q_{\mu}, \hat{\Phi}^a_{\mu}\})    &=& \bar{F}^{\pm}_{w} (\{d\mu_{\mathrm{Haar}},\hat{\Phi}^a_{\omega}\}).\nonumber
\end{eqnarray}
The criterion for the continuous settings takes the form
\begin{equation}
\bar{F}^{-}_{w} (\{d\mu_{\mathrm{Haar}},\hat{\Phi}^a_{\omega}\})<\frac{1}{d^2}.
\end{equation}
The Werner threshold $1/d^2$ has been proven to be a tight bound: If it is achieved, the conditional states should admit a LHS model~\cite{Wiseman1}. In other words,
\begin{equation}
\frac{1-w}{d}<\frac{1}{d^2}
\end{equation}
 is a necessary and sufficient condition for the Werner state to be steerable, while $\frac{1-w}{d}<\mathfrak{F}^-_{\mathrm{NST}}( \{q_{\mu}, \hat{\Phi}^a_{\mu}\})$ is just a sufficient one. Therefore, one can have a state-independent relation,
\begin{equation}
\mathfrak{F}^-_{\mathrm{NST}}( \{q_{\mu}, \hat{\Phi}^a_{\mu}\})\leqslant \frac{1}{d^2},
\end{equation}
where the Werner threshold $1/d^2$ is the upper bound of an arbitrary $\mathfrak{F}^-_{\mathrm{NST}}( \{q_{\mu}, \hat{\Phi}^a_{\mu}\})$.  So, it is reasonable that the criterion in Eq.~\eqref{Wernertype} is referred to as the Werner-type one.

\subsection{ Detecting the steerability of Werner state}
For the qubit case, one can easily verify that $\mathfrak{g}^-_{\mathrm{NST}}=-\mathfrak{g}^+_{\mathrm{NST}}$ and $\bar{f}^{-}=-\bar{f}^+$ for arbitrary measurements $\{q_{\mu}, \hat{\Phi}^{a}_{\mu}\}$. The WJD-type geometric criterion, $ \bar{f}^+>\mathfrak{g}^+_{\mathrm{NST}}$, and the Werner-type one, $\bar{f}^-<\mathfrak{g}^-_{\mathrm{NST}}$, are equivalent. Therefore, only one of the above two criteria, usually the WJD-type one, is required to detect the steerability of the two-qubits  states including the Werner state for $d=2$. This equivalence can also be easily explained since $\mathfrak{F}^+_{\mathrm{NST}}+\mathfrak{F}^-_{\mathrm{NST}}=1$ holds for $d=2$. However, for the high-dimensional system, this equality does not  hold any more. For the Werner state, the maximal value of the averaged fidelity is shown in Eq.~\eqref{avgf}, $\bar{F}^+_{w}(\{q_{\mu}, \hat{\Phi}^a_{\mu}\})=(d-1+w)/[d(d-1)]$. Certainly,  $\bar{F}^+_{w}(\{q_{\mu}, \hat{\Phi}^a_{\mu}\})\leqslant1/(d-1)$. With the WJD bound $H_d/d=(1+1/2+...+1/d)/d$, one can easily check that $\bar{F}^+_{w}(\{q_{\mu}, \hat{\Phi}^a_{\mu}\})<H_d/d$ when $d>2$. Using Eq.~\eqref{NST+5}, one can have
\begin{equation}
\bar{F}^+_{w}(\{q_{\mu}, \hat{\Phi}^a_{\mu}\})< \mathfrak{F}^+_{\mathrm{NST}}( \{q_{\mu}, \hat{\Phi}^a_{\mu}\}),~\mathrm{if} ~d>2.
\end{equation}
So, if $d>2$, the steerability of the Werner state cannot be detected by the WJD-type criterion $\bar{F}^+>\mathfrak{F}^+_{\mathrm{NST}}( \{q_{\mu}, \hat{\Phi}^a_{\mu}\})$.  This is the reason why both  types of the criteria in Eqs.~\eqref{WJDtype} and~\eqref{Wernertype} are required for high-dimensional systems.

\subsection{A criterion with entanglement fidelity}
In the discussions above,  it has been shown that it is usually difficult to calculate the extremal values $\bar{F}^+$ of the fidelity, even for the two-qubit state. With a maximally entangled state $\vert \psi_+ \rangle =\frac{1}{\sqrt{d}}\sum_{k=1}^{d}\vert kk\rangle$, a more convenient criterion may be constructed for the kind of states
\begin{equation}
W_{\varepsilon}=\mathbb{I}_d\otimes \varepsilon(\vert \psi_+\rangle\langle \psi_+\vert),
\end{equation}
which have been used for the discussion about distillation protocols~\cite{Horo}. The entanglement fidelity of $\varepsilon$ is defined as~\cite{Schu}
\begin{equation}
f(\vert\psi_+\rangle, \varepsilon)=\langle\psi_+\vert W_{\varepsilon}\vert\psi_+\rangle,
\end{equation}
which provides a measure of how well the entanglement is preserved by $\varepsilon$. For the state $W_{\varepsilon}$, the measurement $\hat{\Pi}^a_{\omega}$ performed by Alice can be fixed as $(\hat{\Pi}^a_{\omega})^*=\hat{\Phi}^a_{\omega}$, and with $\hat{\Phi}^a_{\omega}=U^{\dagger}_{\omega}\vert a\rangle \langle a\vert U_{\omega}$, one can obtain $\hat{\Pi}^a_{\omega}\otimes\hat{\Phi}^a_{\omega}=(U_{\omega}^*\otimes U_{\omega})^{\dagger}(\hat{P}_{a}\otimes \hat{P}_a)U_{\omega}^*\otimes U_{\omega}$, where $\hat{P}_{a}=\vert a\rangle \langle a\vert$. Now, the averaged fidelity $\bar{F}$ becomes
\begin{equation}
\label{fen}
\bar{F}=\sum_{a=0}^{d-1}\mathrm{Tr}[\hat{P}_{a}\otimes \hat{P}_a\int d\mu_{\mathrm{Harr}}(\omega)U_{\omega}^*\otimes U_{\omega}W_{\varepsilon}(U_{\omega}^*\otimes U_{\omega})^{\dagger}].
\end{equation}
For a Hermitian operator  $\hat{A}$ in a $d$-dimensional Hilbert space, one can define a depolarizing channel $\varepsilon_{\eta}$ as
\begin{equation}
\varepsilon_{\eta}(\hat{A})=\eta\hat{A}+(1-\eta)\mathrm{Tr}(\hat{A})\frac{I_d}{d},
\end{equation}
and the isotropic states in Eq.~\eqref{isotropic} can be expressed as
$W^{\eta}_d=\mathbb{I}_d\otimes \varepsilon_{\eta}(\vert\psi_+\rangle \langle \psi_+\vert)$. As shown in Ref.~\cite{Horo}, an isotropic state can be obtained from $W_{\varepsilon}$ with the twirling procedure
\begin{equation}
W^{\eta}_d=\int d\mu_{\mathrm{Harr}}U_{\omega}^*\otimes U_{\omega}W_{\varepsilon}(U_{\omega}^*\otimes U_{\omega})^{\dagger}.
\end{equation}
Putting this result into Eq.~\eqref{fen}, one can have $\bar{F}=\eta+(1-\eta)/d$. Moreover, with the entanglement fidelity of the depolarizing channel $f(\vert \psi_+\rangle, \varepsilon_{\eta})=\eta+(1-\eta)/d^2$, one can come to $(d+1)\bar{F}=d f(\vert \psi_+\rangle, \varepsilon_{\eta})+1$. With the fact that the entanglement fidelity is invariant under the twirling procedure, say, $f(\vert \psi_+\rangle, \varepsilon)=f(\vert \psi_+\rangle, \varepsilon_{\eta})$, finally,
 \begin{equation}
 \bar{F}=\frac{d f(\vert \psi_+\rangle, \varepsilon)+1}{d+1}.
 \end{equation}
Now, the criterion $\bar{F}> H_d/d$ can be reexpressed as
\begin{equation}
\label{fc}
f(\vert \psi_+\rangle, \varepsilon)>\frac{1}{d}\left(\frac{d+1}{d}H_d-1\right),
\end{equation}
which is a sufficient condition for $W_{\varepsilon}$ to be steerable from A to B.

Here, we say that a channel $\varepsilon$ is entanglement preserving (EP) if it is not an EB channel. It is shown in Appendix~\ref{appB} that a sufficient condition for the EP channel is
\begin{equation}
f(\vert \psi_+\rangle, \varepsilon)>\frac{1}{d}.
\end{equation}
Noting that $[(d+1)H_d/d-1]/d>1/d$, one may also apply the criterion in~Eq.~\eqref{fc} as a sufficient condition for $\varepsilon$ to be an  EP channel. However, not every EP channel can be applied for constructing a steerable $W_{\varepsilon}$.  For example, when
\begin{equation}
\frac{1}{d}<f(\vert \psi_+\rangle, \varepsilon_{\eta})\leqslant \frac{1}{d}\left(\frac{d+1}{d}H_d-1\right),
\end{equation}
the depolarizing channel is EP but the isotropic state is un-steerable from A to B.

\section{Conclusions}
\label{Sec6}

According to the fundamental idea that a steering inequality can be constructed by just considering the measurements performed by Bob,  proposed in Refs.~\cite{can22,sau,Joness}, and from the definitions of steering from Alice to Bob \cite{Can1}, we have developed a general scheme for designing linear steering criteria for high-dimensional systems. For a given set of measurements (on Bob's side), we have defined two  quantities, the so-called nonsteering thresholds. If the measured averaged fidelity exceeds these thresholds, the state shared by Alice and Bob is steerable from Alice to Bob, and the measurements performed by Alice are also verified to be incompatible. Within the general scheme, we also constructed a LSI when the set of measurements performed by Bob has a continuous setting. In the derivation of this LSI, the results in Refs~\cite{Wiseman1,Werner} have been applied. Two kinds  of criteria, the WJD type  and Werner type, can be applied as the sufficient conditions of steerability for bipartite state.  For the qubit case, it has been shown that the two types of steering criteria are equivalent to each other. However, when $d>2$, these criteria have different properties.

The LSI in this work is limited for the case where the set of measurements by Bob has a continuous and equal-weighted form. From the view of experiment, the LSIs with a finite number of experimental settings are required. We leave such kinds of LSIs, especially adapted to the Werner state, as our future works. We expect that the results in this work could lead to further theoretical or experimental consequences.

\acknowledgements
This work was supported by the National Natural Science Foundation of China under Grants No.~12047576 and No.~11947404.

\appendix

\section{Proof of Eq.~\eqref{decoposition}}
\label{appA}

First, a convenient tool is required to be introduced, where a bounded operator in a $d$-dimensional Hilbert space $\mathcal{H}_{d}$ is related to a vector in an enlarged Hilbert space $\mathcal{H}_{d}^{\otimes 2}$. Let $A$ be a bounded operator on $\mathcal{H}_d$, where $A_{ij}=\langle i\vert  A\vert j\rangle$ are the matrix elements and $\{|i\ket\}_{i=1}^d$ is a fixed basis. An isomorphism between $A$ and a vector $\vert A\rangle\rangle$ in $\mathcal{H}_{d}^{\otimes 2}$ can be defined
\begin{equation}
\label{isomorphism}
\vert A\rangle\rangle =\sqrt{d} A\otimes I_d\vert \psi_+\rangle=\sum_{i,j=1}^d A_{ij}\vert ij\rangle,
 \end{equation}
where $\vert ij\rangle=\vert i\rangle\otimes \vert j\rangle$, and $\vert \psi_+ \rangle =\frac{1}{\sqrt{d}}\sum_{k=1}^{d}\vert kk\rangle$ is a maximally entangled state in $\mathcal{H}_{d}^{\otimes 2}$. This isomorphism offers a one-to-one mapping between a matrix and its vector form. For three arbitrary bounded operators $A$, $B$, and $\rho$ on $\mathcal{H}_{d}$, it is easy to verify
\be
\bra\bra A|ij\ket&=&\bra j|A^\dag|i\ket,\ \ \bra ij|A\ket\ket=\bra i|A|j\ket,\\
\tr[A^{\dagger}B]&=&\langle\langle A\vert B\rangle\rangle,\ \ 
\vert A\rho B\rangle\rangle =A\otimes B^{\mathrm{T}}\vert \rho \rangle\rangle,
\ee
where $\langle\langle A\vert=\sum_{i,j=1}^d A^*_{ij}\langle ij\vert$, $A^\dag$ is the adjoint of $A$, and $B^{\mathrm{T}}$ is the transpose of $B$.

Now, a $d^2\times d^2$ density operator $W$ can be expressed as
\begin{equation}
W=\sum_{m=1}^{d^2}\lambda_m\vert\Psi_m\rangle\langle \Psi_m\vert,\nonumber
\end{equation}
where $\vert\Psi_m\rangle$ are the normalized eigenvectors of the density operator $W$,  and $\lambda_m$ are the corresponding eigenvalues, $\sum_{m}^{d^2}\lambda_m=1$. According to the isomorphism in Eq.~\eqref{isomorphism}, for each vector $\vert \Psi_m\rangle$, there is a corresponding matrix $\Gamma_m$ satisfying $\vert \Psi_m\rangle =\vert \Gamma_m\rangle\rangle$, and therefore, the density matrix $W$ can be also expressed as
\begin{equation}
W=\sum_{m}\lambda_m\vert \Gamma_m\rangle\rangle\langle\langle \Gamma_m\vert.\nonumber
\end{equation}
Furthermore, for Alice's reduced density operator $\rho_{\mathrm{A}}=\tr_{\mathrm{B}}W$, one can have
\begin{eqnarray}
\langle i\vert \rho_{\mathrm{A}}\vert j\rangle
&=&\sum_{k=1}^{d}\langle ik\vert \sum_{m=1}^{d^2} \lambda_m \vert \Gamma_m\rangle\rangle\langle\langle \Gamma_m\vert jk\rangle\nonumber\\
&=&\sum_{m=1}^{d^2}\sum_{k=1}^{d}\lambda_m \langle i\vert\Gamma_m\vert k\rangle\langle k\vert \Gamma_m^{\dagger}\vert j\rangle\nonumber\\
&=&\langle i\vert \sum_m\lambda_m\Gamma_m\Gamma_m^{\dagger}\vert j\rangle, \nonumber
\end{eqnarray}
which means
\begin{equation}
\rho_{\mathrm{A}}^{\mathrm{T}}=\sum_m\lambda_m\Gamma_m^{*}\Gamma_m^{\mathrm{T}}.\nonumber
\end{equation}

With denotations introduced above, it will be shown that the density operator $W$ can be rewritten as 
\begin{equation}
\label{dW}
W=\mathbb{I}_d\otimes \varepsilon(\vert \sqrt{\rho_A}\rangle\rangle\langle\langle \sqrt{\rho_A}\vert)
\end{equation}
in the following two cases.

(a) In the case that the reduced density operator $\rho_\mathrm{A}$ is full-rank, $\det \rho_\mathrm{A}\neq 0$, and the operator $\left(\sqrt{\rho_\mathrm{A}^{\mathrm{T}}}\right)^{-1}$ is well-defined. A set of Kraus operators
\begin{equation}
\label{B_m}
B_m=\sqrt{\lambda_m}\Gamma^{\mathrm{T}}_m\left(\sqrt{\rho_\mathrm{A}^{\mathrm{T}}}\right)^{-1},
\end{equation}
can be constructed to represent a linear map $\varepsilon(\rho)=\sum_m B_m\rho B_m^\dagger$, where $\rho$ is a density operator on $\mathcal{H}_d$. One may check that $\varepsilon$ is trace preserving by verifying $\sum_mB_m^{\dagger}B_m=I_d$, which can be proved as follows:
\be
\sum_mB_m^{\dagger}B_m&=&\sum_m\lambda_m\left(\sqrt{\rho_\mathrm{A}^{\mathrm{T}}}\right)^{-1}\Gamma_m^*\Gamma^{\mathrm{T}}_m\left(\sqrt{\rho_\mathrm{A}^{\mathrm{T}}}\right)^{-1}\non\\
&=&\left(\sqrt{\rho_\mathrm{A}^{\mathrm{T}}}\right)^{-1}\rho_\mathrm{A}^{\mathrm{T}}\left(\sqrt{\rho_\mathrm{A}^{\mathrm{T}}}\right)^{-1}\non\\
&=&I_d.
\ee
Finally, it is an easy task to show that 
\be
W&=&\sum_{m}\lambda_m\vert \Gamma_m\rangle\rangle\langle\langle \Gamma_m\vert\non\\
&=&\sum_m\vert \sqrt{\rho_\mathrm{A}}B_m^{\mathrm{T}}\rangle\rangle \langle \langle \sqrt{\rho_\mathrm{A}}B_m^{\mathrm{T}}\vert\non\\
&=&\mathbb{I}_d\otimes \varepsilon(\vert \sqrt{\rho_\mathrm{A}}\rangle\rangle\langle\langle \sqrt{\rho_\mathrm{A}}\vert).
\ee

(b) In the case that $\mathrm{det}\rho_\mathrm{A}=0$, the operator $\left(\sqrt{\rho_\mathrm{A}^{\mathrm{T}}}\right)^{-1}$ is not well-defined in the Hilbert space $\mathcal{H}_d$. Denote the rank of $\rho_\mathrm{A}$ by $d'$, and $\mathcal{H}_d$ can be decomposed as $\mathcal{H}_d=\mathcal{H}_{d'}\oplus \mathcal{H}_{\bar{d}}$, where $\mathcal{H}_{d'}$ is the support of $\rho_\mathrm{A}$, and $d'+\bar{d}=d$. Now, the operator $\left(\sqrt{\rho_\mathrm{A}^{\mathrm{T}}}\right)^{-1}$ can be well-defined in the subspace $\mathcal{H}_{d'}$, and Eq.~\eqref{B_m}
can still be employed to obtain $B_m$ in $\mathcal{H}_{d'}$, and $\sum_m B_m^{\dagger}B_m=I_{d'}$. Besides the $\{B_m\}$ defined above, it is required to introduce another set of operators $\{\bar{B}_n\}$ in $\mathcal{H}_{\bar{d}}$ satisfying
\begin{equation}
\label{barb}
\sum_{n}\bar{B}_n^{\dagger}\bar{B}_n=I_{\bar{d}},
\end{equation}
where  $I_{\bar{d}}$ is the identity operator on the subspace $\mathcal{H}_{\bar{d}}$. Then, a superoperator $\varepsilon$ can still be defined as
\be
\varepsilon(\rho)=\sum_m B_m\rho B_m^{\dagger}+\sum_n \bar{B}_n\rho \bar{B}_n^{\dagger},
\ee
with $\sum_m B_m^{\dagger}B_m+\sum_n \bar{B}_n^{\dagger}\bar{B}_n=I_d$.  Similarly with the proof for case (a), the density operator $W$ can also be expressed as
\be
W&=&\sum_{m}\lambda_m\vert \Gamma_m\rangle\rangle\langle\langle \Gamma_m\vert\non\\
&=&\sum_m\vert \sqrt{\rho_A}B_m^{\mathrm{T}}\rangle\rangle \langle \langle \sqrt{\rho_A}B_m^{\mathrm{T}}\vert+\sum_n\vert \sqrt{\rho_A}\bar{B}_n^{\mathrm{T}}\rangle\rangle \langle \langle \sqrt{\rho_A}\bar{B}_n^{\mathrm{T}}\vert\non\\
&=&\mathbb{I}_d\otimes \varepsilon(\vert \sqrt{\rho_A}\rangle\rangle\langle\langle \sqrt{\rho_A}\vert),
\ee
where $\sqrt{\rho_A}\bar{B}_n^{\mathrm{T}}=0$ has been used.

From the discussions above, it can be seen that, in principle, every bipartite state can be decomposed as the form in Eq.~\eqref{decoposition} by choosing $|\Psi\ket=|\sqrt{\rho_A}\ket\ket$. Moreover, it should be noted that the decomposition in Eq.~\eqref{decoposition} is not unique. For example, with a given decomposition $W=\mathbb{I}_d\otimes\varepsilon( \vert\Psi\rangle\langle\Psi\vert)$, a new state $\vert\tilde{\Psi}\rangle=I_d\otimes U\vert\Psi\rangle$ can be introduced with $U$ a local unitary transformation, and meanwhile, a new superoperator $\tilde{\varepsilon}=\varepsilon\circ \mathcal{U}^{\dagger}$ can be defined with $\mathcal{U}^{\dagger}$ a unitary channel $\mathcal{U}^{\dagger}(\rho)=U^\dag\rho U$. Therefore, $\tilde{\varepsilon}(\rho)=\varepsilon(U^{\dagger}\rho U)$, and the density operator $W$ can be reexpressed as 
\be
W=\mathbb{I}_d\otimes \tilde{\varepsilon}( \vert\tilde{\Psi}\rangle\langle\tilde{\Psi}\vert),
\ee
which shows that the state $\vert\Psi\rangle$ and the linear map $\varepsilon$ in Eq.~\eqref{decoposition} are two tightly related objects.

Furthermore, for the case (b) where $\rho_\mathrm{A}$ is not full-rank in $\mathcal{H}_d$, construction of $\bar{B}_n$ in Eq.~\eqref{barb} is also not unique. Actually, for $\det\rho_\mathrm{A}=0$, there is a one-to-many relation between the density operator $W$, and the decompositions as in Eq.~\eqref{decoposition} is not unique, even though $\vert\Psi\rangle=\vert \sqrt{\rho_A}\rangle\rangle$ is fixed. In practice, one may just select out one of all the possible channels satisfying Eq.~\eqref{barb} to obtain a decomposition in Eq.~\eqref{decoposition}.
\qed

\section{Entanglement fidelity of a EB channel}
\label{appB}

For a set of Kraus operators $\{B_m\}_{m=1}^{d^2}$ of a quantum channel $\varepsilon$, one may introduced the so-called process matrix $\hat{\lambda}_{\varepsilon}$~\cite{Xu}
\begin{equation}
\hat{\lambda}_{\varepsilon}=\sum_m B_m\otimes B_m^*.
\end{equation}
In the vector form, the input state $\rho$ of the channel $\varepsilon$ and the output state $\varepsilon(\rho)$ are simply related by
\begin{equation}
\vert \varepsilon(\rho)\rangle\rangle=\hat{\lambda}_{\varepsilon}\vert \rho\rangle\rangle.
\end{equation}
Meanwhile, the trace preserving condition $\sum_mB_m^{\dagger}B_m=I_d$ can be expressed as
\begin{equation}
\label{CPTP}
\langle\langle  I_d\vert=\langle\langle I_d\vert \hat{\lambda}_{\varepsilon}.
\end{equation}
The entanglement fidelity $f(\vert\psi_+\rangle, \varepsilon)$ becomes
\begin{equation}
f(\vert\psi_+\rangle, \varepsilon)=\frac{1}{d^2}\tr \hat{\lambda}_{\varepsilon}.
\end{equation}

According to Ref.~\cite{Namiki4}, any EB channel can be described by a positive-operator-valued measurement (POVM) $\{M_y\}$ with $M_y^{\dagger}=M_y$, where $y$ denotes an outcome occurring with a probability $\tr(\rho M_y)$ and a reconstruction rule $y\rightarrow \rho_y$ determines that $\rho_y$ is prepared when the measurement outcome is $y$. Then, the channel acts as
\begin{equation}
\varepsilon_\mathrm{EB}(\rho)=\sum_y\rho_y\mathrm{Tr[}\rho M_y].
\end{equation}
Based on this, it can be verified that the assemblage $\{\tilde{\rho}^a_{\mu}\}$ in Eq.~\eqref{unnormalized} resulted from the EB channel always admits a LHS decomposition.  The above equation can be rewritten as $\vert\varepsilon_\mathrm{EB}(\rho)\rangle\rangle=\sum_y\vert \rho_y \rangle\rangle  \langle\langle M_y\vert \rho\rangle\rangle$, and the process matrix for the EB channel is
\begin{equation}
\hat{\lambda}_{\varepsilon_\mathrm{EB}}=\sum_y\vert \rho_y \rangle\rangle  \langle\langle M_y\vert.
\end{equation}
Certainly, the trace preserving condition in Eq.~\eqref{CPTP} is satisfied. The entanglement fidelity $f(\vert\psi_+\rangle,\varepsilon_\mathrm{EB})$ can be calculated
\be
f(\vert\psi_+\rangle, \varepsilon_\mathrm{EB})=\frac{1}{d^2}\tr\hat{\lambda}_{\varepsilon_\mathrm{EB}}=\frac{1}{d^2}\sum_y\tr(\rho_yM_y).
\ee
With $\mathrm{Tr}(\rho_yM_y)\leqslant\tr M_y$ and $\sum_y\tr M_y=d$, one can find
\begin{equation}
f(\vert\psi_+\rangle,
\varepsilon_\mathrm{EB})\leqslant\frac{1}{d}.
\end{equation}
For a depolarizing channel $\varepsilon_{\eta}$, the condition $f(\vert\psi_+\rangle,\varepsilon_{\eta})>1/d$ is necessary and sufficient for the isotropic states to be entangled~\cite{Horo}.

\end{document}